\documentclass[draft]{agujournal2019}
\usepackage{url}
\usepackage{lineno}
\usepackage[inline]{trackchanges}
\usepackage{soul}

\draftfalse

\journalname{JGR: Space Physics}

\begin{document}

\title{Energetic proton dropouts during the Juno flyby of Europa strongly depend on magnetic field perturbations.}

\authors{H.L.F. Huybrighs\affil{1}, S. Cervantes\affil{1}, P. Kollmann\affil{2}, C. Paranicas\affil{2}, C.F. Bowers\affil{1}, X. Cao\affil{1}, M.K.G. Holmberg\affil{1}, C.M. Jackman\affil{1}, S. Brophy Lee\affil{1}, A. Bl\"{o}cker\affil{3}, E. Marchisio\affil{1,4}}

\affiliation{1}{Astronomy \& Astrophysics Section, School of Cosmic Physics, Dublin Institute for Advanced Studies, DIAS Dunsink Observatory, Dublin D15 XR2R, Ireland}
\affiliation{2}{Johns Hopkins Applied Physics Laboratory, Laurel, MD, USA}
\affiliation{3}{Space Research Institute, Austrian Academy of Sciences, Schmiedlstrasse 6, 8042 Graz, Austria}
\affiliation{4}{School of Physical Sciences, Dublin City University, Glasnevin, D09 W6Y4, Ireland}

\correspondingauthor{Hans Huybrighs}{hans@cp.dias.ie}

\begin{keypoints}
\item Magnetic field perturbations due to moon-magnetosphere interaction drive 100 keV-1 MeV energetic proton losses in Europa's wake.
\item At $\sim$1 MeV proton losses also occur due to the proton's short half bounce periods between Jupiter's magnetic poles.
\item The pitch angle distribution of the proton dropout is very sensitive to the specific modeled magnetic field perturbations.

\end{keypoints}

\begin{abstract}
During Juno’s only flyby of Europa, the Jupiter Energetic Particle Detector Instrument (JEDI) measured complex dropouts in the energetic ion flux in Europa’s wake. We investigate the causes of these dropouts, focusing specifically on energetic protons of $\sim100$ keV and $\sim1$ MeV, using back-tracking particle simulations, a prescribed description of Europa's atmosphere and a three-dimensional single fluid magnetohydrodynamics (MHD) model of the plasma-atmosphere interaction.

We investigate the role of magnetic field perturbations resulting from the interaction between Jupiter's magnetospheric plasma and Europa's atmosphere and the presence of field-aligned electron beams in Europa’s wake. We compare the simulated effect of the perturbed fields on the pitch angle distributions of the ion losses to Juno-JEDI measurements. We find that at $\sim100$ keV, field perturbations are the dominant factor controlling the distribution of the losses along the flyby, while at $\sim1$ MeV a combination of field perturbations and absorption by the surface due to short half bounce periods is required to explain the measured losses. We also find that the effect of charge-exchange with Europa's tenuous atmosphere is weak and absorption by dust in Europa's environment is negligible.

Furthermore, we find that the perturbed magnetic fields which best represent the measurements are those that account for the plasma interaction with a sub-/anti-Jovian asymmetric atmosphere, non-uniform ionization of the atmosphere, and electron beams. This sensitivity to the specific field perturbation demonstrates that combining observations and modeling of proton depletions constitute an important tool to probe the electromagnetic field and atmospheric configurations of Europa.

\end{abstract}

\section*{Plain Language Summary}
Jupiter’s moon Europa has a potentially habitable ocean under its icy surface. It is located in Jupiter's magnetosphere and interacts with the magnetic field and plasma from that magnetosphere. Measurements of the fields and plasma are crucial in determining the habitable conditions of its ocean. The Juno spacecraft measured localized decreases of magnetospheric energetic protons in Europa's plasma wake behind the moon. We investigate the cause of these dropouts by comparing the measurements of Juno with simulations of the protons. 

We find that the losses of 100 keV protons are mainly caused by complex perturbations in the magnetic field in Europa’s wake. One case of perturbed fields describes the measurements best. In that case the perturbations are the result of the interaction of plasma with Europa's asymmetric atmosphere, non-uniform ionization of the atmosphere and beams of electrons moving along the magnetic field lines. Because the proton losses vary for different magnetic field perturbations, we can use the proton measurements to study the magnetic field's structure. At 1 MeV energies, the losses are more complex as half bounce motion of the energetic protons between Jupiter's magnetic poles becomes an important loss factor too. 

\section{Introduction}
Jupiter's moon Europa orbits within Jupiter's magnetosphere, where it interacts with the surrounding magnetic fields and plasma. Highly energetic ions ($>10$s keV) from Jupiter's magnetosphere interact with Europa's surface and atmosphere. The dynamics of these ions are influenced by perturbed magnetic fields near Europa.

Prior to Juno, the Galileo spacecraft encountered Europa's wake during flybys E4, E11 and E15. Flyby E17 also encountered the wake, but not within Europa's downstream cross section. Energetic ion dropouts were reported during these encounters \cite{Paranicas2000EnergeticEuropa,Pappalardo2009_Kivelson}. These previous studies reported a strong energy and pitch angle dependence of the ion losses, which occurred in a region in the wake roughly the size of Europa. While they did not reach a conclusion on the cause of the dropouts, various hypotheses were provided. These loss mechanisms included proton dropouts due to the size of their gyroradius exceeding the orbital altitude or losses associated with ions that have short half bounce times compared to their azimuthal drift across the moon.

Previously, using a combination of energetic ion tracing simulations and analysis of Galileo Energetic Particle Detector (EPD) data, \citeA{Huybrighs2020,Huybrighs2021,Huybrighs2023} argued that the energetic proton dropouts near Europa are caused by a combination of different effects. These are (1) impact of the energetic ions on the surface, (2) deflection of the protons by perturbed electromagnetic fields resulting from the interaction of the magnetospheric plasma with Europa's tenuous atmosphere and ionosphere, (3) charge exchange of the protons with atmospheric neutrals converting energetic protons into energetic neutral atoms (ENAs), (4) and potentially water plumes via proton deflection by perturbed fields and charge exchange with the plumes. \citeA{Huybrighs2020,Huybrighs2021,Huybrighs2023} considered one Galileo flyby upstream of Europa (E26) and two remote flybys downstream (E17 and E25A).

On 29 September (DOY 272) 2022, the Juno mission passed through Europa's wake (from 09:33:48 to 09:36:47 UTC) and approached the moon's surface as close as 355 km. During this encounter Juno made in-situ measurements with multiple instruments to characterize Europa's moon-magnetosphere interaction. These instruments detected Europa's ionosphere \cite{Parisi2023}, perturbed plasma flows \cite{Ebert2025}, pickup ion species including H$^+$ and O$_{2}^+$ \cite{Szalay2024,Ebert2025}, field aligned electron beams with energies from 30 to 300 eV \cite{Allegrini2024}, perturbed magnetic fields \cite{Addison2024,Cervantes2025}, plasma waves \cite{Kurth2023}, a tentative dust detection \cite{Kurth2023}, energetic electron dropouts $>1$ keV \cite{Paranicas2023} \& $>15$ MeV \cite{Herceg2024} and, of particular importance here, dropouts in energetic ions ($>10$s keV) compared to the upstream magnetospheric population  \cite{Clark2025}.

\citeA{Clark2025} identified energetic ion dropouts in Europa's wake using measurements by the Jupiter Energetic Particle Detector Instrument (JEDI) \cite{Mauk2017TheMission}, which can measure species resolved energy and angular distributions. Specifically, JEDI distinguishes among hydrogen, oxygen, and sulfur ions over an energy range of $\sim50$ keV – 4 MeV, $\sim300$ keV – 14 MeV, $\sim400$ keV – 15 MeV, respectively. For protons, S$^{n+}$ and O$^{n+}$, depletions centered around 90$^{\circ}$ pitch angle are visible from at least several $10$s keV for protons and several $100$s keV for S$^{n+}$ and O$^{n+}$. Specifically for protons, the dropout widens in pitch angle range from $\sim200$ keV and becomes asymmetrical with respect to 90$^{\circ}$.
Regarding the measured dropouts concentrated near 90$^{\circ}$ pitch angle, \citeA{Clark2025} states that 'ion loss via absorption to the moon is favored for particle pitch angles where their mirror point is near the magnetic or centrifugal equator, therefore the ion has a high probability of intersecting the moon along its drift path'.

In this study we trace particles backwards in time from the JEDI detectors and investigate flux decreases at the detector for different scenarios. We will show that for a case without perturbed fields, the effect proposed in \citeA{Clark2025}, where 90$^{\circ}$ pitch angle particles intersect Europa's surface, cannot explain the dropout structure  in itself. While we do not dismiss this effect entirely, our simulations show that the intersection effect is minor. Instead, using a magnetohydrodynamics (MHD) model, which provides for a more realistic description of the electromagnetic field near Europa, we show that inclusion of realistic fields is very critical to understanding the size and shape of the flux decrease in the plasma wake region for 100 keV and 1 MeV protons. At energies of 100 keV, perturbed fields are the most important consideration required to reproduce the losses. However, our model is not able to fully reproduce the losses at 1 MeV. We propose that, at these energies, the observed loss results from a combination of perturbed fields and a shortening of the ion half bounce period. As the half bounce period becomes sufficiently short relative to the ions' azimuthal drift across the moon, absorption by Europa's surface leads to dropouts in the wake. We also highlight that while charge exchange with the atmosphere could play a role, its effect is likely weak during this flyby. The role of dust is fully negligible, as we will argue. 

This new study combining modelling and data distinguishes itself from our previous work because the proton dropouts are investigated in a new physical regime: close to Europa in its downstream geometrical wake (i.e. inside the downstream cross section of Europa) where magnetic field perturbations typical for this region are present and at energies where losses due to short half bounce periods could occur. Furthermore, the capability of Juno's instruments allow for major advancements. Firstly, the time, energy and pitch angle resolution of the Juno-JEDI data is far higher than during the Galileo E17 and E25A flybys, allowing a more conclusive interpretation of the measurements. Secondly, measurements of Juno-JADE allowed for the first detection of field aligned electron beams in Europa's wake \cite{Allegrini2024}, which affect the proton dropouts, as we will show. This is the first study that investigates the effect of the field-aligned electron beams on the energetic proton fluxes. Various other studies using either data analysis or particle modeling have also investigated the role of surface absorption, perturbed fields and charge exchange on the ion fluxes e.g. \citeA{Paranicas2000,Paranicas2007,Breer2019,Addison2021,Addison2022,Nordheim2022}.

\section{Methodology}
In this section we first describe the Juno JEDI data used for this study (Section \ref{s_data}). Then we describe the particle tracing simulations that we use to compare with the measurements (Section \ref{s_tracing}). The last section presents the MHD simulations used to model the perturbed electromagnetic fields near Europa, which is crucial to accurately simulate the proton trajectories (Section \ref{s_mhd}).

\subsection{Juno JEDI data}
\label{s_data}
\begin{figure}
\begin{center}
\noindent\includegraphics[width=1.0\textwidth]{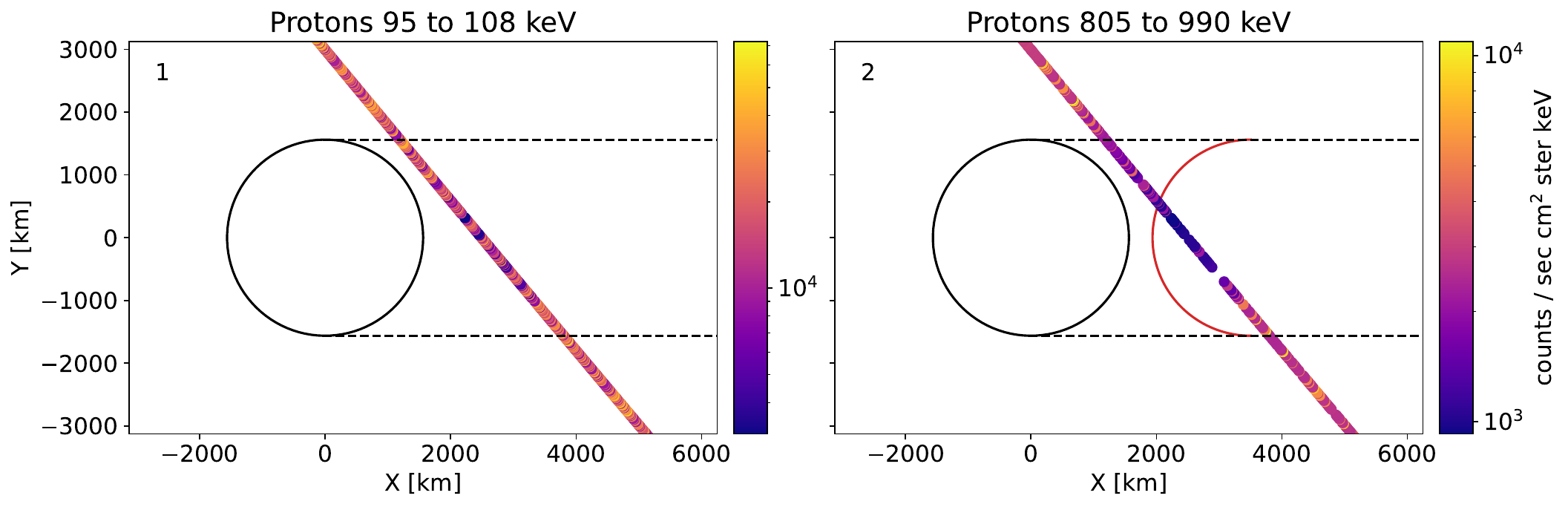}
\end{center}
\caption{Juno's flyby geometry colored by the measured flux in two energy channels. A dropout region is visible in the wake at both energies. The dashed line indicates the location of Europa's wake. The red semi circle in panel 2 (right) represents the radius of Europa shifted along the x-axis by the azimuthal drift distance, during half a bounce period, of a 1 MeV particle with a 89 $^{\circ}$ pitch angle assuming 50 $\mathrm{km~s^{-1}}$ azimuthal drift (see also Figure \ref{fig_bounce_period}). The section of Juno's flyby that lies to the right of the red semi circle (and is within the wake), is an approximation of the region where dropouts due to short half bounce periods are most likely to occur. Coordinate system: EPhiO, z-axis is parallel to Jupiter’s spin axis, y-axis points towards Jupiter, and x-axis along the direction of the corotational plasma flow. For reproducibility we specify that data in panel 1 (left) correspond to JEDI channel T01PF07, panel 2 (right) to channel T01PF20.}
\label{fig_traj}
\end{figure}

Figure \ref{fig_traj} shows an overview of the Juno flyby geometry with the flux of energetic protons overplotted.
We use the EPhiO coordinate system throughout this paper, in which the z-axis is parallel to Jupiter’s spin axis, the y-axis points towards Jupiter, and the x-axis completes the right-handed system along the direction of the corotational plasma flow. We define the entry and exit time of the wake when Juno's x-coordinates correspond to Europa's radius (1560.8km), i.e. where the trajectory projected on the z=0 plane intersects the dotted lines in Figure \ref{fig_traj}.
These limits differ slightly from the definition of the geometrical wake in e.g. \citeA{Szalay2024}. The limits in our study approximately represent the range along the flyby over which proton dropouts due to surface absorption could occur, which corresponds to Europa's size in the z=0 plane.

In this paper we compare data and simulations at energies of $\sim100$ keV (specifically 95.11-108.387 keV) and $\sim1$ MeV (specifically 805.88-990.44 keV). Note that for simplicity we will not repeat the exact range throughout the paper. We focus exclusively on proton simulations and data from JEDI's proton (mass separated) channels. We use the proton channels because they extend to lower energy than the heavy ion channels. The individual heavy ion channels (e.g., for S$^{n+}$ and O$^{n+}$) on JEDI start at higher energies and there is further uncertainty since JEDI does not discriminate among charge states \cite{Clark2016ChargeMagnetosphere}, which affects the size of gyroradius. We focus on the two energy channels mentioned above to examine the difference in measured dropout signatures and to investigate the potential role of charge exchange. The charge exchange cross section decreases strongly with increasing incoming proton energy. At $\sim100$ keV the charge exchange cross section is relatively large and charge exchange is more likely to occur \cite{Huybrighs2020,Huybrighs2023}, whereas at $\sim1$ MeV the probability of charge exchange occurring is negligible due to the small cross section \cite{Huybrighs2021,Huybrighs2023}. Here we use an extrapolated cross section based on \citeA{Basu1987LinearObservations}, see Appendix A in \citeA{Huybrighs2023}.

To compare the JEDI data with our particle tracing simulations, we will represent the Juno JEDI proton measurements as time-versus-pitch angle spectra. These spectra were made by combining the proton data from all the available directional channels for the energy ranges 95.11-108.387 keV or 805.88-990.44 keV. The proton data are from sensor JEDI-270 only, as for this flyby there are no data available for JEDI-90, the other sensor with proton separation capabilities. Count rates in individual channels are low, therefore binning in pitch angle (3.6$^{\circ}$ bins) and time (24 second bins) has been applied.

\subsection{Particle tracing}
\label{s_tracing}

\begin{table}
\caption{Parameter values used in the particle tracing simulations.}
\centering
\begin{tabular}{l c}
\hline
 Technical parameters particle tracing simulation  & Value  \\
\hline
  Integration time step particle trajectory [s]  & 0.001  \\
  Number of integration time steps per gyration  & $\sim150$ \\
  Maximum number of time steps per particle &  100000\\
  Number of particles per spacecraft trajectory step &  10000  \\
  Number of energy bins & 10 \\
  Spacecraft trajectory time step [s] & 2 \\
  Magnetic field for non-perturbed case [nT] & (77.2, -120.2, -422.0) \\
  \hline
\label{tab_particle}
\end{tabular}
\end{table}

We simulate the measured flux of energetic protons using a Monte Carlo particle back-tracing code, developed from \citeA{Huybrighs2017OnMission,Huybrighs2020,Huybrighs2021,Huybrighs2023,Huybrighs2024} and previously applied to interpret energetic ion measurements from the Galileo mission near Europa and Io. Particle tracing has been employed previously at various moons to investigate the ion dynamics as affected by field perturbations and atmospheric charge exchange (e.g. \citeA{Selesnick2009ChargeIo,Kotova2015ModelingDione,Poppe2018ThermalMagnetosphere,Breer2019,Liuzzo2019EnergeticCallistob,Plainaki2020,Addison2021InfluenceWeathering,Nordheim2022,Liuzzo2024}).

Here we provide a short overview of the simulation setup. The particle tracing model and loss processes are described in further detail in \citeA{Huybrighs2023}. The relevant input parameters are summarized in Table \ref{tab_particle}. 

Because the energetic protons have negligible influence on the electromagnetic fields they can be treated as test particles. In the simulation, protons are traced back in time from the Juno spacecraft's position. For each back traced proton, it is determined if it impacts and/or charge exchanges. If the simulated proton impacts Europa's surface or charge exchange with the atmosphere occurs at any point along the trajectory, the particle is considered lost. This analysis is performed for many particles to determine the depletion as a fraction of the (normalized) undisturbed flux. 

\begin{figure}
\begin{center}
\noindent\includegraphics[width=1.0\textwidth]{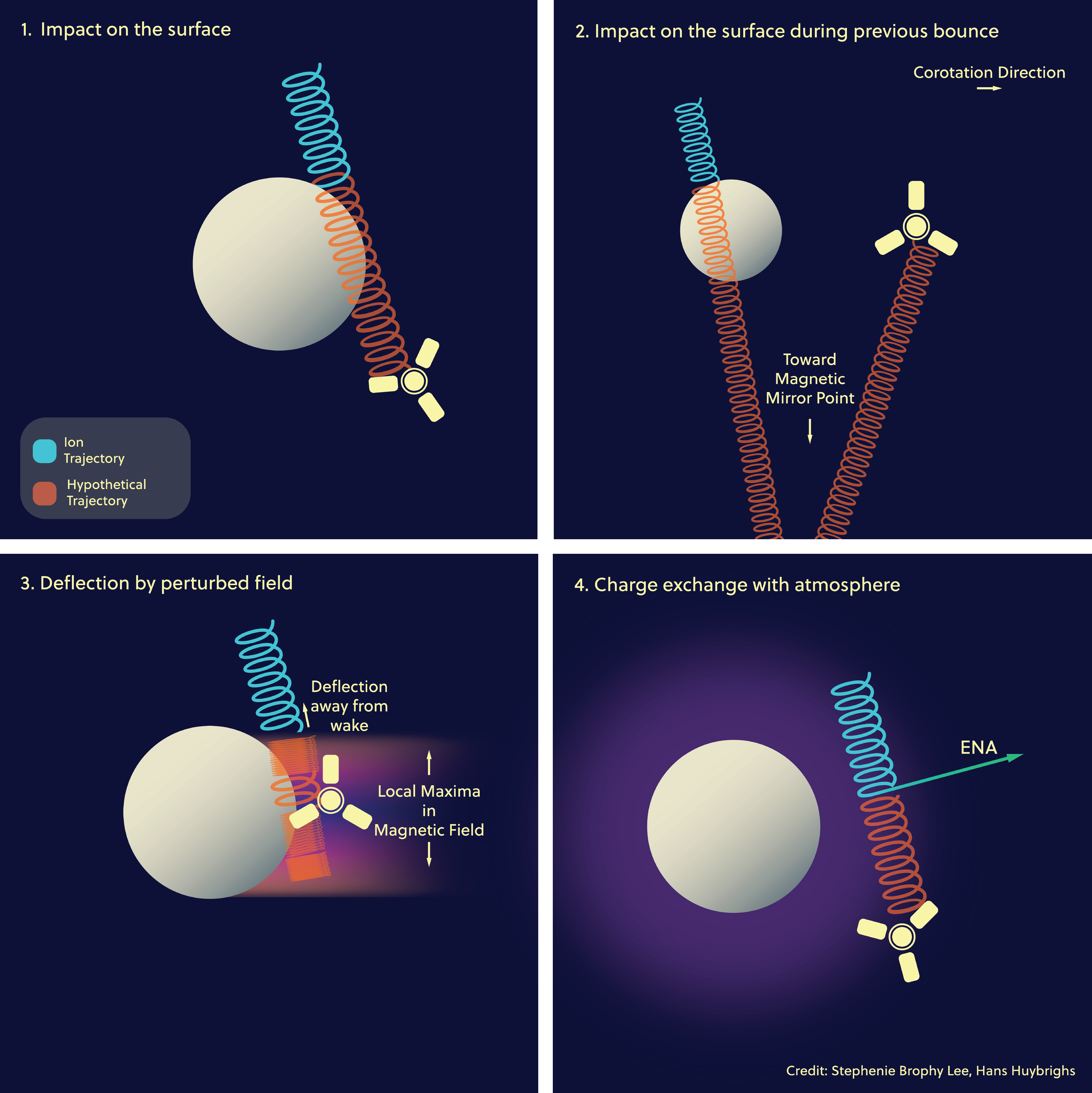}
\end{center}
\caption{Four key loss processes of energetic protons considered in this study. Particles prior to the loss are colored blue. Red indicates hypothetical trajectories corresponding to the case where the particle continues existing after the loss. Panel 1: Impact on the surface occurs when the particle trajectory intersects with Europa's surface before it reaches the detector. Panel 2: Losses due to impact on the surface also occur downstream of Europa if the half bounce period is short enough that depleted protons will leave a gap in the proton distribution during their next bounce motion(s). Panel 3: Perturbed fields deflect protons from certain regions. In the back tracing simulation this is represented by particles that mirror locally and impact on Europa's surface due to perturbed fields (red). Please note that in case 3 we don't simulate the original deflected particle (blue), so the trajectory shown is only for illustrative purposes. Panel 4: Charge exchange turns the energetic proton into an Energetic Neutral Atom (ENA), leading the particle to escape from Europa's environment instead of being detected.}
\label{fig_sketch}
\end{figure}

In the simulation three types of losses of the back traced particles can occur, with some important nuances in their interpretation. A schematic illustrating these loss processes is shown in Figure \ref{fig_sketch}.
Firstly, if a particle that is traced back in time impacts on Europa's surface, it must originate from the surface in forward time. Thus, the particle cannot exist, as energetic protons do not originate from Europa's surface. An important nuance occurs in this interpretation if perturbed fields are also present, as we will discuss in the next paragraph. Secondly, if a particle charge exchanges in the back tracing simulation it means that in forward time the energetic proton would have to originate from Europa's tenuous atmosphere, which is not considered a possible source of energetic protons. Therefore, it is considered that the particle has been lost to charge exchange and could not have been detected by JEDI. 

Thirdly, losses can also occur when the perturbed fields are included in the particle tracing simulation. In this case the back traced protons are altered, for example their gyroradius can be changed or they could deflect due to the gradient drift associated with local maxima in the electromagnetic field. With perturbed fields, back traced protons can impact on the surface of Europa, where they would miss it in the non perturbed case. However, this does not necessarily mean that, in forward time, the perturbed fields enhance the losses to the surface. A second hypothesis is that, in forward time, the particle does not interact with the surface and has actually been deflected away by local maxima in the perturbed magnetic fields, from the JEDI position and viewing angle from which the particle is traced back in time. We suggest that the losses occurring due to the perturbed fields are a combination of losses to the surface and deflection away from that region, however the back tracing simulations don't explicitly distinguish between these two. Note that the regions in which losses occur, which are the result of back traced protons impacting Europa's surface due to perturbed fields, are referred to as 'forbidden regions'.

We simulate the trajectories of energetic protons under different scenarios, including those with perturbed (inhomogeneous), not-perturbed (homogeneous) electromagnetic fields, and with and without charge exchange with Europa's tenuous atmosphere. In the non-perturbed case, we neglect any perturbations in the electromagnetic field resulting from the interaction of the corotational plasma with Europa's atmosphere or its induced dipole. The components of the non-perturbed field are listed in Table \ref{tab_particle}. While they are slightly different from the initial conditions in the MHD simulations, the difference is not important for our conclusions. While the situation with non-perturbed fields is not expected to occur at Europa in reality, including this case allows us to determine the relative contribution of the perturbed fields to the proton losses. In this work we will evaluate different perturbed magnetic fields and their relative contribution to the proton dropout.
We will consider perturbations resulting from the plasma interaction with a sub-/anti-Jovian asymmetric atmosphere, non-uniform ionization of the atmosphere and electrons beams, as discussed in more detail in Section \ref{s_mhd}.

\begin{figure}
\begin{center}
\noindent\includegraphics[width=1.0\textwidth]{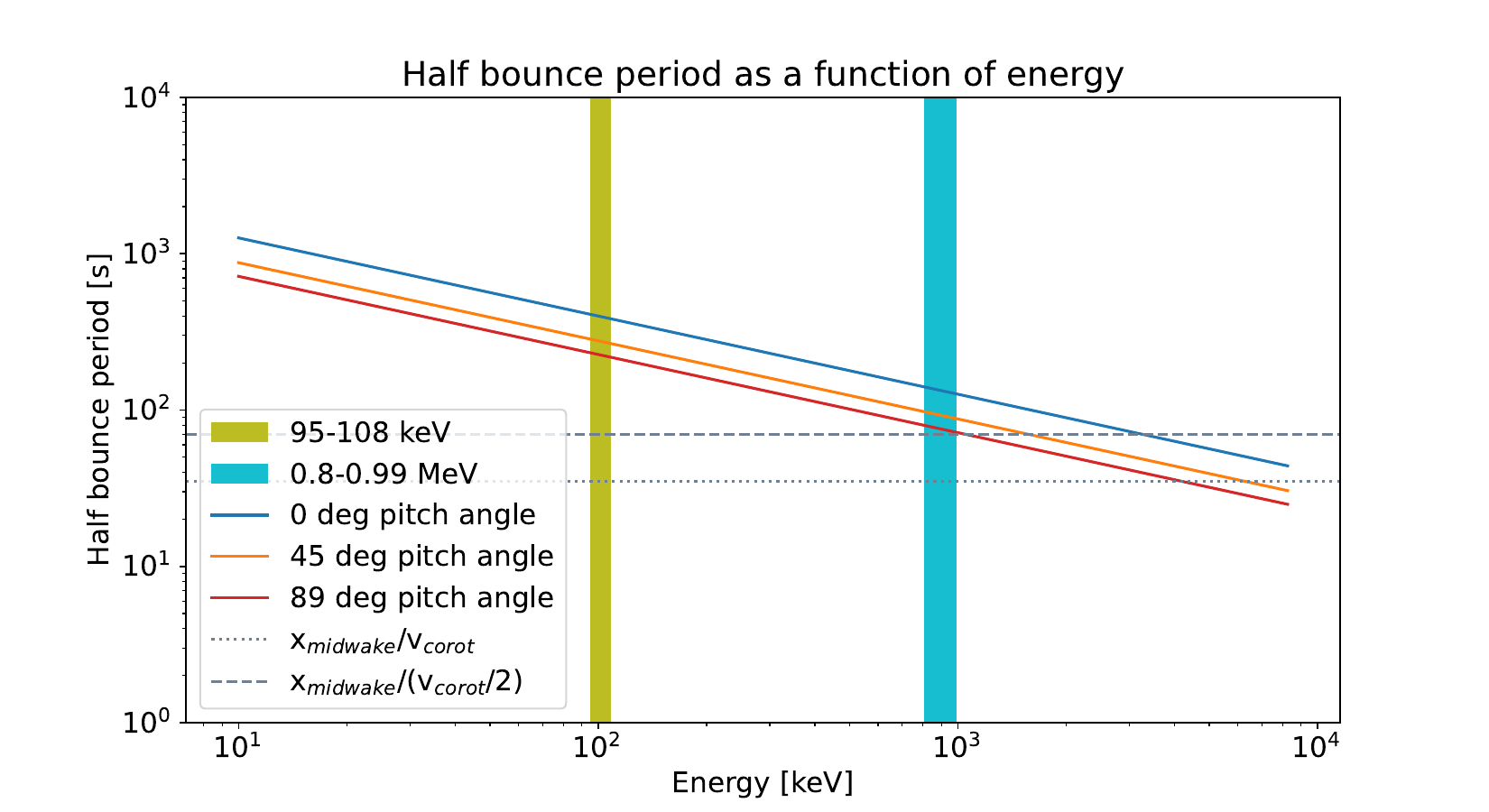}
\end{center}
\caption{Half bounce period of energetic protons calculated using Equation \ref{eq_bounce}, based on Equation 3.14 from \citeA{Baumjohann1997}. The horizontal lines indicate the time needed for a particle to drift along the corotation direction from Europa to Juno's x position in the middle of the wake ($x_{midwake}$). Where the half bounce period lines intersect the horizontal lines we can expect losses to propagate in the wake due to depleted particles failing to re-enter the wake in their second bounce motion. We consider two horizontal lines, reflecting undisturbed corotation speed (v$_{corot}$) of 100 $\mathrm{km~s^{-1}}$ and slow down by 50\%.}
\label{fig_bounce_period}
\end{figure}

For protons of $\sim100$ keV the half bounce periods (see Figure \ref{fig_bounce_period}) are expected to be larger than the time it takes the particle to drift azimuthally over the distance between Europa and Juno. 
However at $\sim1$ MeV the half bounce periods become shorter and it becomes possible for a proton to drift azimuthally within Juno's distance to Europa. Therefore we could expect additional losses to occur due to the short half bounce periods of these protons. This effect is not accounted for in our simulation, because protons are assumed not to be able to enter the simulation box during a second bounce motion. We will show that this causes a discrepancy between the modelled and measured flux at energies of $\sim1$ MeV. Proton half bounce periods $(\tau_b)$ for energies relevant to this study are shown in Figure \ref{fig_bounce_period}. These half bounce periods were calculated using Equation \ref{eq_bounce}, which is based on Equation 3.14 from \citeA{Baumjohann1997}. In this equation, $L$ is the L-shell corresponding to Europa's position, R$_J $ is the radius of Jupiter (71492 km), $W$ the particle energy in eV, $m$ the particle mass in kg, and $\alpha_{eq}$ the equatorial pitch angle of the particle. In Figure \ref{fig_traj} the azimuthal drift distance occuring during a bounce period is also represented.

\begin{equation}
    \tau_b \approx \frac{LR_J}{(W/m)^{(1/2)}}(3.7-1.6 \sin \alpha_{eq})
    \label{eq_bounce}
\end{equation}

In Europa's environment the energetic proton flux decreases with increasing proton energy. Thus, in the two JEDI energy channels we consider ($\sim$ 100 keV and $\sim$ 1 MeV), more protons will be detected corresponding to the lower energy limit of the channel than the higher limit. The variation of flux in each of the channels is less than an order of magnitude \cite{Clark2025}. To take into account this decrease in flux in the particle simulation, we have assumed a change in flux of, respectively, $\sim$10 percent and $\sim$25 over the energy channels. We note that we found in \citeA{Huybrighs2020} that the simulated proton losses are not very sensitive to the specific flux decrease in one energy channel. In the simulation, the two Juno energy bins we consider are divided in ten energy bins. This resolution is assumed to be more than sufficient as the gyroradius changes less than 11\% over these (narrow) channels.

\subsection{MHD simulations}
\label{s_mhd}

The perturbed electromagnetic fields that arise due to the interaction between Europa's atmosphere and the ambient magnetospheric plasma can influence the trajectories of energetic protons. In this study, we obtain these inhomogeneous fields using the three-dimensional (3D) single-fluid MHD simulations of \citeA{Cervantes2025}, who applied a modified version of the publicly available PLUTO code \cite{Mignone2007} and adjusted it to the conditions of the Juno flyby at Europa. The PLUTO code has already been successfully employed in the context of moon-magnetosphere interaction by \citeA{Duling2022} and \citeA{Strack2024} in order to model Ganymede's and Callisto's plasma interaction, respectively.

\citeA{Cervantes2025} presents a detailed description of the MHD simulations setup, numerics and input parameters; here we only provide a summary of the key points relevant to this study. The model uses a spherical grid centered on Europa and describes Europa's plasma interaction with the following variables: magnetic field $\underline{B}$, plasma bulk velocity $\underline{v}$, plasma mass density $\rho$, and total thermal pressure $p$, by solving a set of four MHD evolution equations consisting of the continuity equation, the momentum equation, the induction equation, and an equation for the internal energy.  It includes collisions between ions and neutrals, plasma production due to electron impact ionization and photoionization, loss due to dissociative recombination between ions and neutrals, and electromagnetic induction in a subsurface water ocean. A summary of the parameters for the initial and boundary conditions in this model is given in Table~\ref{tab_mhd}. 

One of the main findings of the Juno flyby at Europa was the discovery of bi-directional magnetic field-aligned electron beams with energies from $\sim$30 to 300 eV downstream of the moon \cite{Allegrini2024}. \citeA{Cervantes2025} included the beams in their MHD simulations as sheets of locally enhanced electron impact ionization, and their results show that the beams fill the wake with freshly ionized plasma. In addition, \citeA{Cervantes2025} studied the effect of the $\mathrm{O_2}$ distribution around Europa on the moon's plasma environment by considering two atmospheric descriptions: a radially symmetric atmosphere, and a sub-/anti-Jovian asymmetric atmosphere with maximum (minimum) column density in the anti-Jovian (sub-Jovian) apex. The properties of both descriptions are summarized in Table~\ref{tab_mhd}. Furthermore, \citeA{Cervantes2025} also investigated the effects of two impact ionization scenarios caused by background magnetospheric electrons: (1) a constant uniform ionization frequency across the entire simulation domain and (2) a non-uniform ionization frequency driven by decrease in electron temperature within Europa's atmosphere wake, as modeled by \citeA{Saur1998InteractionAtmosphere}.  The values of the background electron impact ionization frequencies for both cases are presented in Table~\ref{tab_mhd}.

\begin{table}
\caption{Initial and boundary conditions, atmospheric properties, and background electron impact ionization frequencies used in the MHD model}
\centering
\begin{tabular}{l c}
\hline
 Initial and boundary condition values in the EPhiO system & Value  \\
 \hline
  $\underline{B}_0$ [nT] & (70, -120, -420)  \\
  $\underline{v}_0$ [$\mathrm{km~s^{-1}}$] & (100, 0, 0)  \\
  $\rho_0$ [$\mathrm{amu~m^{-3}}$] & $1.85\times10^9$  \\
  $p_0$ [nPa] & 4.8 \\ 
  \hline
 Atmospheric properties  & Value  \\
 \hline
  Average mass of neutral particles [amu] & 32 \\
  Scale height\textsuperscript{ a} [km] & 100 \\
  Surface number density: radially symmetric atmosphere [$\mathrm{m^{-3}}$] & $10^{13}$ \\
  Surface number density: sub-/anti-Jovian asymmetric atmosphere [$\mathrm{m^{-3}}$] & $5\times10^{12} - 10^{13}$ \\
  Column density: radially symmetric atmosphere [$\mathrm{m^{-2}}$] & $10^{18}$ \\
  Column density: sub-/anti-Jovian asymmetric atmosphere [$\mathrm{m^{-2}}$] & $5\times10^{17} - 10^{18}$ \\
  \hline
 Background electron impact ionization frequencies & Value \\
  \hline
 Uniform case: constant everywhere [$\mathrm{s^{-1}}$] & $4.6\times10^{-6}$ \\
 Non-uniform case: upstream [$\mathrm{s^{-1}}$] & $4.6\times10^{-6}$ \\
 Non-uniform case: center of the wake [$\mathrm{s^{-1}}$] & $2\times10^{-6}$ \\
 Non-uniform case: flanks of the wake [$\mathrm{s^{-1}}$] & $10^{-6}$ \\
  \hline
  \multicolumn{1}{l}{\textsuperscript{a }\footnotesize{The scale height value is the same in both the symmetric and the asymmetric descriptions}.}
   \\
\hline
\label{tab_mhd}
\end{tabular}
\end{table}

\citeA{Cervantes2025} performed five MHD runs and followed a stepwise approach in order to study the effect of each of the above-mentioned parameters on Europa's plasma interaction. The details of the simulations are shown in Table~\ref{tab_mhd runs}. Their initial, most simple run (Run 1) accounts for a radially symmetric atmosphere and a uniform ionization by the background magnetospheric electrons and neglects the presence of the electron beams. In subsequent runs, they included an asymmetric atmosphere (Run 2), a non-uniform background ionization around Europa (Run 3), and a series of electron beams in the moon's wake (Run 5). The authors show that the setup of the latter provides the best fit to the magnetic field and plasma density measurements from Juno. Note that we do not consider Run 4 from \citeA{Cervantes2025} in this study, since its setup is very similar to the one in Run 5.

In the particle tracing simulations the MHD fields are interpolated on a resolution of 0.05 R$_E$ ($\sim 80$km). We consider this a reasonable approach, as field perturbations of smaller scale are not produced in the MHD simulations. However, in reality perturbations occur on the scale of several $10$s of km in Europa's wake \cite{Cervantes2025}. To date models of the magnetic field have not been able to reproduce these smaller scale perturbations and their cause is not understood. The $\sim 80$km scale is comparable or smaller in size than the gyoradius of 100 keV protons or $\sim 1$ MeV protons of pitch angle $\textless 30^\circ$ or $\textgreater150^\circ$. This means that these protons could experience additional magnetic field magnitude (Grad-B) drift due to the small scale perturbations. However, since our model is able to reproduce the general properties of the measured proton dropout feature at 100 keV (where the protons would be most sensitive to this scale of perturbations), we assume that the smaller scale perturbations only have a minor effect on the dropout structure. When models become available that can reproduce the smaller scale magnetic field perturbations, new particle tracing studies could be conducted for a detailed investigation of the role of smaller scale perturbations on the energetic protons. 

\begin{table}
\caption{Summary of MHD Runs from \citeA{Cervantes2025}}
\centering
\begin{tabular}{c c c c}
\hline
Run & $\mathrm{O_2}$ atmosphere & Background ionization frequency & Electron beams  \\ \hline
1 & Radially symmetric & Uniform & No  \\
2 & Sub-/anti-Jovian asymmetric & Uniform & No  \\
3 & Sub-/anti-Jovian asymmetric & Non-uniform & No  \\
5 & Sub-/anti-Jovian asymmetric & Non-uniform & Yes \\ 
\hline
\label{tab_mhd runs}
\end{tabular}
\end{table}

\section{Results and discussion}
In Section \ref{ss_sims}, we show simulations of the dropout with and without perturbed fields and compare them to the data. We use the perturbed fields from 'Run 5' (see Section \ref{s_mhd}), unless otherwise mentioned. We show that the field perturbations are crucial for explaining the morphology of the pitch angle distribution of the dropout, especially near 100 keV. Then, in Section \ref{ss_1Mev}, we demonstrate that short half bounce periods become important in explaining the dropout structure for proton energies near 1 MeV, in addition to the perturbed fields. In Section \ref{ss_sims_fields}, we also show that the dropout structure is sensitive to different magnetic field perturbations, by comparing simulation results for proton energies near 100 keV under the perturbed fields defined by Run 1 to 3 to those defined in Run 5. We argue that Run 5 describes the data the best. Then, in Section \ref{ss_charge_exchange}, we discuss the weak contribution of charge exchange with Europa's tenuous atmosphere to the proton losses. Finally, in Section \ref{ss_dust}, we argue that dust detected in Europa's wake has a negligible role on the proton dropouts. 

\subsection{Field perturbations are crucial for explaining the measured dropout structure}
\label{ss_sims}
\begin{figure}
\begin{center}
\noindent\includegraphics[width=1.0\textwidth]{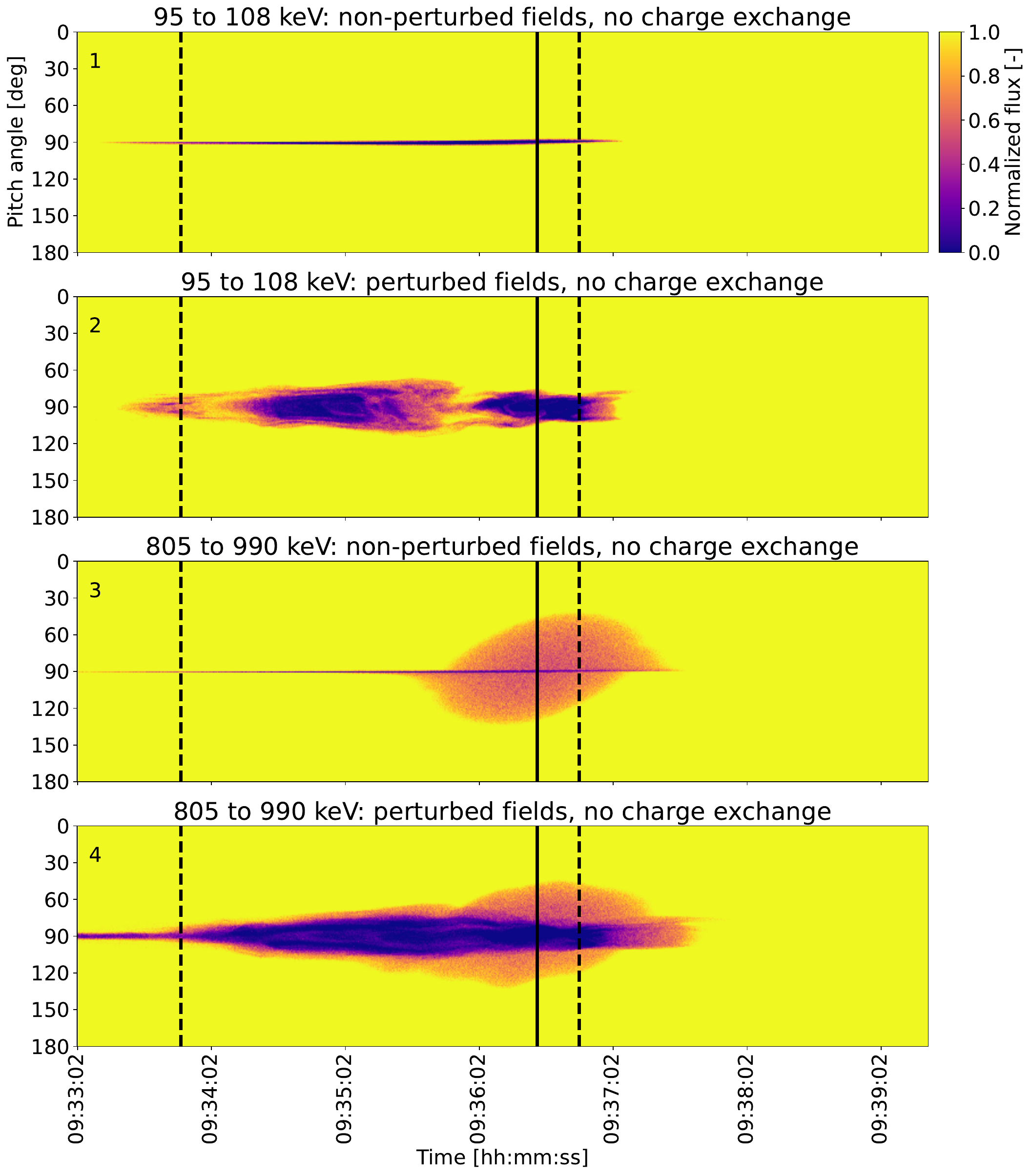}
\end{center}
\caption{Panels 1 and 2 show the simulated dropout for $\sim$ 100 keV as a function of time and pitch angle, with and without perturbed fields (Run 5 from \citeA{Cervantes2025}). Panels 3-4 show the same for $\sim$ 1 MeV. In the non-perturbed case the depletion is very narrow, while in the perturbed case, the depletion structure is wider and has complex structure. The solid vertical line represents the closest approach, the dashed vertical lines indicate the entry and exit of Europa's wake. Pitch angle values of zero are at the top of the panel as field aligned particles would be coming from the 'north' of Europa (positive z).}
\label{fig_sims}
\end{figure}

Our simulations reveal a significant difference in the simulated dropout at both the lower ($\sim100$ keV) and higher ($\sim1$ MeV) energy channels between the case with and without perturbed fields (see Figure \ref{fig_sims}). At $\sim100$ keV under non-perturbed field conditions, the dropout is confined to close to 90$^{\circ}$ pitch angle. This dropout results from proton trajectories intersecting with Europa's surface (as previously discussed in \citeA{Huybrighs2023}). Effectively, protons that mirror very close to Europa's magnetic latitude would be absent in the plasma wake region. However, particles with different mirror latitudes can easily escape Europa absorption. In the perturbed fields case, the dropout becomes wider in pitch angle and attains a complex structure. At $\sim$ 1 MeV we also observe a wider and more complex depletion structure throughout the wake. At $\sim$ 1 MeV, the dropout has an asymmetrical structure with respect to the 90$^{\circ}$ pitch angle near the closest approach. This asymmetry strongly depends on the tilt of the background magnetic field, which determines the access of energetic protons to Europa's surface; reducing the tilt would reduce the asymmetry.

\begin{figure}
\begin{center}
\noindent\includegraphics[width=1.0\textwidth]{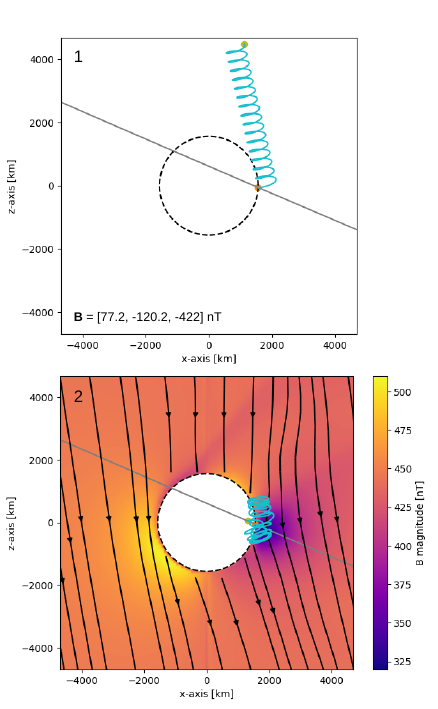}
\end{center}
\caption{Figure of a $\sim$ 1 MeV particle trajectory traced back in time from Juno's position (orange dot), without perturbed fields (panel 1) and with perturbed fields (panel 2, fields from Run 5 from \citeA{Cervantes2025}). The start position (orange dot) is the same in both panels. The final point (green) is different. When the perturbed fields are included, ions that would not have collided under non-perturbed field conditions with the moon do encounter it, creating a localized dropout in flux. The grey solid line indicates the spacecraft trajectory.}
\label{fig_particle}
\end{figure}

The cause of the widening is demonstrated in Figure \ref{fig_particle} which shows a proton back traced from Juno JEDI under conditions with and without perturbed fields. Our particle simulations show that when the perturbed fields are included, back traced ions that would not have collided with the moon under non-perturbed field conditions do encounter it, creating a localized dropout in flux. Due to gradients in the perturbed fields, the particle mirrors locally and impacts on the surface of Europa. As discussed in Section \ref{s_tracing}, in forward time, this would mean the particle has either impacted on the surface or has been deflected away from the wake. The widening of the pitch angle dropout and the underlying cause are consistent with the effects previously considered in \citeA{Huybrighs2023}, which analyzed two remote downstream flybys that did not pass through the geometrical wake.

\begin{figure}
\begin{center}
\noindent\includegraphics[width=1.0\textwidth]{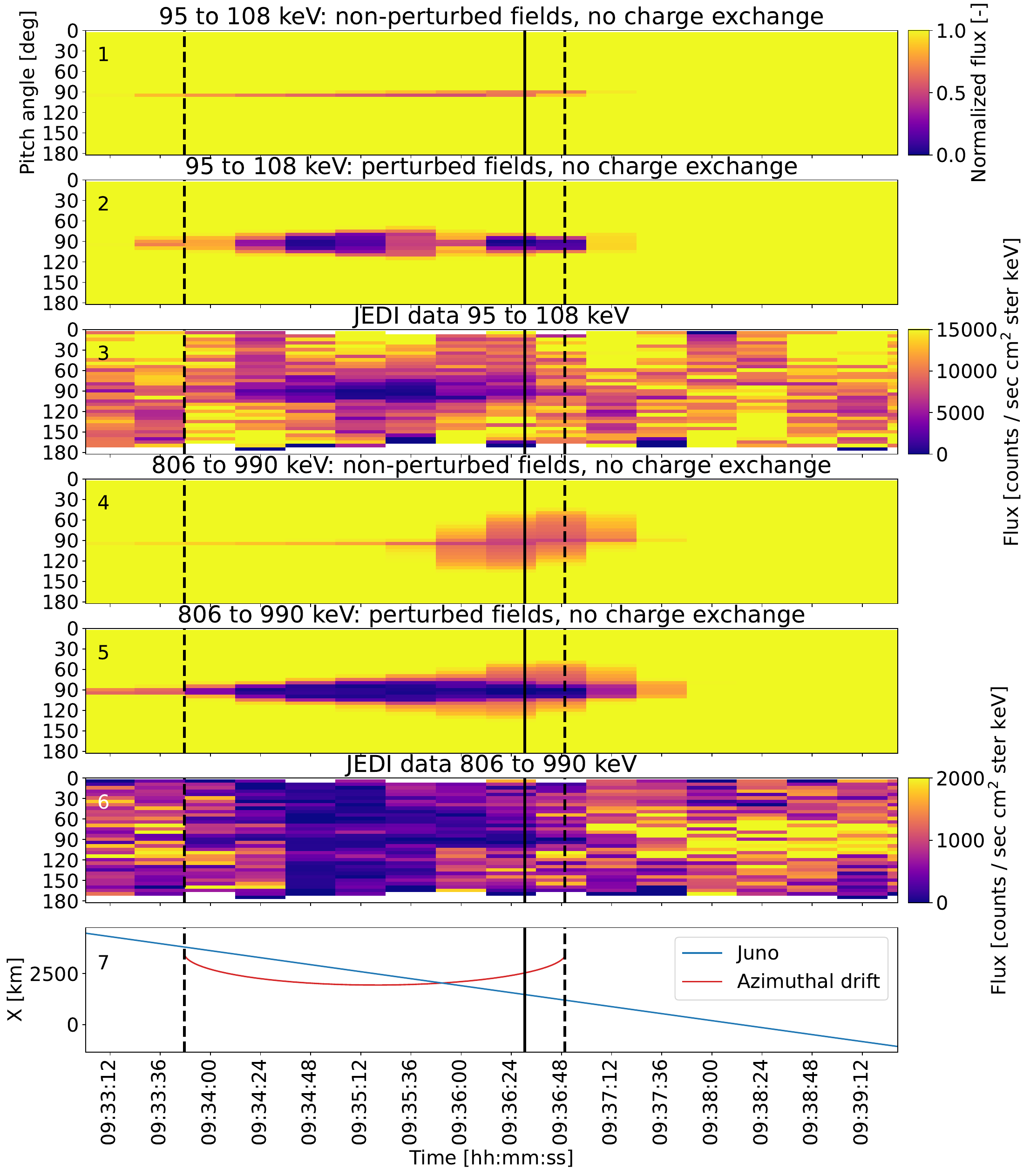}
\end{center}
\caption{In this figure the simulations from Figure \ref{fig_sims} (panel 1-2 and 4-5) are compared to the Juno JEDI data (panel 3 and \change{5}{6}). The simulations have been re-binned in pitch angle and time the same way as the data. Note that the colorbar in the data panels (3 and 6) has been saturated at respectively 15000 and 2000. The blue line in the bottom panel shows the x-coordinate of Juno in the EPhiO frame and the red line the
azimuthal drift distance, during a half bounce period, of a 1 MeV proton with a 89$^{\circ}$ pitch angle (see also Figure \ref{fig_traj} and Figure \ref{fig_bounce_period}). The segment of the Juno trajectory (blue line) that lies above the red line is where losses due to short half bounce periods are most likely to occur. The solid vertical line represents the closest approach, the dashed vertical lines indicate the entry and exit of Europa's wake. Note that the colorbar for all simulations, panels 1-2 and 4-5, is the same.}
\label{fig_sims_data}
\end{figure}

In Figure \ref{fig_sims_data} we compare the simulations to the Juno JEDI measurements. The simulations have been re-binned in the same time and pitch angle bins as the data to allow for a direct comparison between the two. Regarding the data at $\sim100$ keV, there is a good qualitative agreement between the length and width of the dropout feature with the simulation that accounts for the perturbed fields. The simulation with the non-perturbed fields fails to account for the width of the dropout feature. This comparison indicates that the perturbed fields are a dominating cause of the dropout at these energies. Charge exchange with the atmosphere could also play a role at these energies as we will discuss in Section \ref{ss_charge_exchange}. 

\subsection{At $\sim1$ MeV losses occur due to short half bounce periods}
\label{ss_1Mev}

At $\sim1$ MeV we see a number of important differences between the modeled dropouts and JEDI observations (Figure \ref{fig_sims_data}). While both the models with perturbed and non-perturbed fields predict that all depletions occur within roughly 50 and 130$^{\circ}$ pitch angle, JEDI observations indicate that a depletion over all pitch angles occurs between 09:34:48 and 09:35:12. We attribute these wide pitch angle losses to the short half bounce period of these protons. The half bounce period at these distances is sufficiently short that particles impacting Europa's surface should have reached the wake on the next bounce, within a distance that is as near to the moon as the closest approach of the flyby.
This is illustrated further in the bottom panel of Figure \ref{fig_sims_data}, which shows the distance of Juno along the x-axis in the EPhiO frame as well the azimuthal drift distance of a 1 MeV proton during half a bounce period with a 89$^{\circ}$ pitch angle and 50 $\mathrm{km~s^{-1}}$ azimuthal drift speed (see also Figure \ref{fig_traj} and \ref{fig_bounce_period}). The assumed slow down corresponds to a conservative reduction of 50\% (see paragraph below). The segment of the Juno trajectory that lies above the red line is where losses due to short half bounce periods are most likely to occur. This explains why the broad pitch angle loss is disappearing from around 09:36:00 as Juno is now closer to Europa than the azimuthal drift occurring during a half bounce period. Around 09:34:00, the broad dropout feature is also absent, as near the entry of the wake protons would only encounter a small cross section of Europa along the corotation (x-axis) direction and the probability of losses occurring is lower. Thus, from 09:36:00 onward, losses are less likely and less frequent due to the short half bounce period. From 09:36:00, observed losses are more concentrated near 90$^{\circ}$ and agree more with the dropout morphology predicted in the simulation accounting for the perturbed fields. In conclusion, we propose that the observed loss feature at $\sim100$ keV is caused by the perturbed fields, while at $\sim1$ MeV the feature is a combination of the short half bounce period (in the middle wake encounter) and of the perturbed fields (towards the exit of the wake encounter).

For estimating the azimuthal drift distance, we assumed an azimuthal drift velocity of 50 $\mathrm{km~s^{-1}}$, which implies a deceleration of the plasma flow by 50$\%$ compared to upstream conditions. While this is a rough approximation, a 50$\%$ decrease in velocity is approximately consistent with the modeling results by \citeA{Cervantes2025} and Juno measurements of the plasma \cite{Szalay2024,Ebert2025}. In fact as shown in \citeA{Cervantes2025}, along the plasma flow streamlines associated with the electron beams, the flow is slowed down even more to 10$\%$ of its upstream value.

\citeA{Clark2025} states that above $\sim200$ keV, there appears to be a strong pitch angle asymmetry in the protons that occurs with respect to the 90$^{\circ}$ pitch angle. This asymmetry is also visible in the $\sim1$ MeV data shown in panel 6 of Figure \ref{fig_sims_data}, near 09:36:00 and 09:36:34. \citeA{Clark2025} argues that the asymmetry could be related to Europa's magnetic latitude offset of the magnetic equator and the resulting longer or shorter half bounce periods of the protons depending on their position above or below Europa in magnetic latitude. While our model is not able to reproduce the dropout due to short half bounce periods, we do argue that this asymmetry is not likely caused by local interactions with Europa, in particular with the perturbed fields, as our model with perturbed fields does not produce similar asymmetric features. A future study that accounts for the short half bounce period of the protons and the complex location-dependent deceleration (e.g. near the electron beams) of the corotating plasma over Europa could be used to further investigate this hypothesis and to constrain the azimuthal drift speed of the protons over Europa. While not in the scope of this study, our results also indicate that future studies should investigate the role of the field perturbations associated with the electron beams on the surface precipitation of the energetic ions.

\subsection{The dropout structure is sensitive to different magnetic field perturbations}
\label{ss_sims_fields}

\begin{figure}
\begin{center}
\noindent\includegraphics[width=1.0\textwidth]{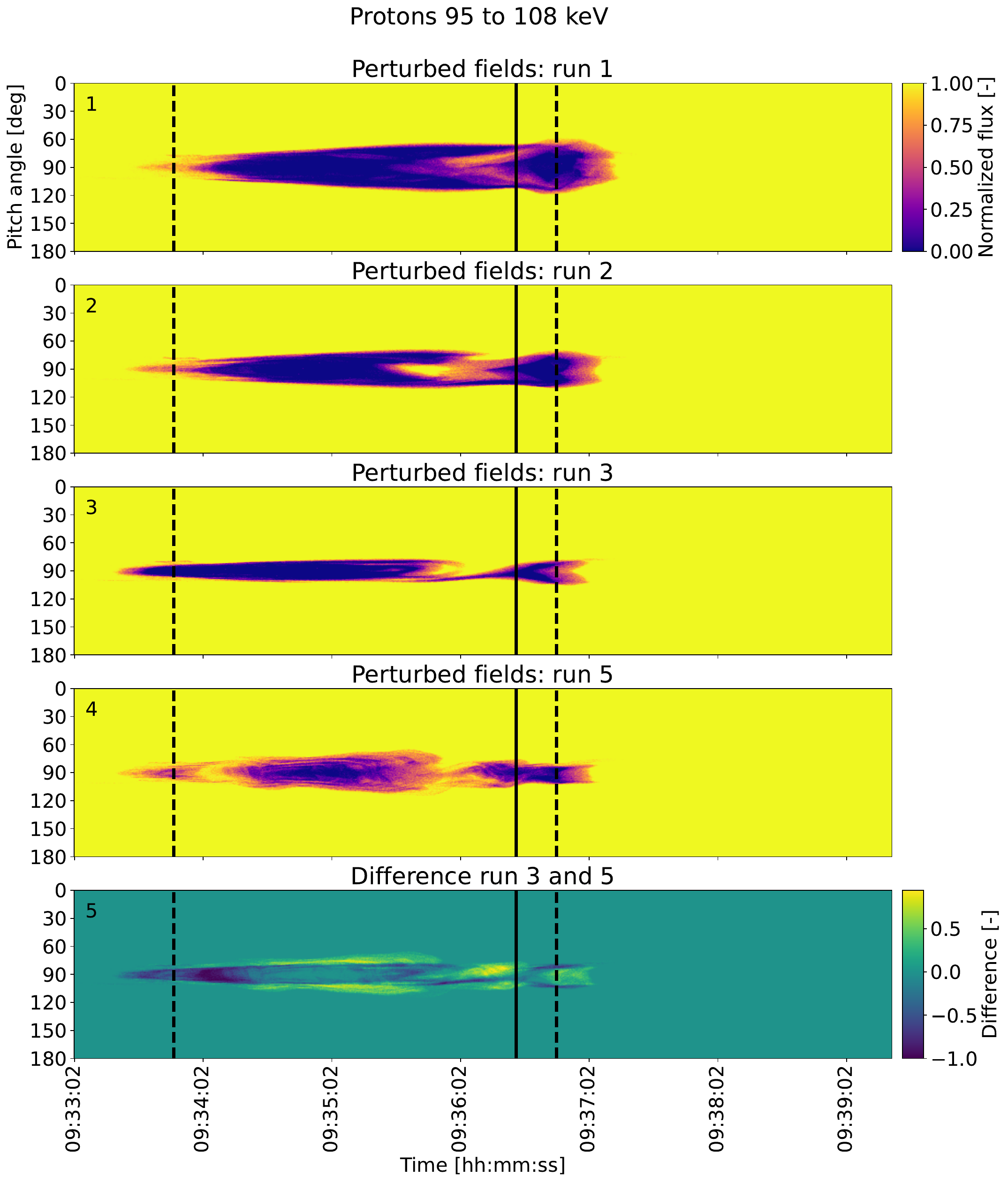}
\end{center}
\caption{Panels 1 to 4: simulated $\sim$ 100 keV proton flux for different magnetic fields from \citeA{Cervantes2025}. For each case a different proton dropout structure is predicted, which implies that the proton dropout structure is sensitive to the specific description of Europa's environment. The bottom panel highlights the importance of including the electron beams (run 5) by showing the quantitative difference between run 3 and run 5. The solid vertical line represents the closest approach, the dashed vertical lines indicate the entry and exit of Europa's wake.}
\label{fig_sims_fields}
\end{figure}

\citeA{Cervantes2025} considers a stepwise approach in building up their MHD model, by starting from a simple case and subsequently increasing its physical complexity (see Section \ref{s_mhd}). An overview of the model runs is shown in Table \ref{tab_mhd runs}.
We have ran the particle tracing code for each of the magnetospheric fields corresponding to Run 1 to Run 3 and Run 5. Figure \ref{fig_sims_fields} shows that the fields of each simulation correspond to a different proton depletion pattern in the simulations. The bottom panel shows the difference between Run 3 (no electron beams) and Run 5 (with electron beams) and highlights the importance of the electron beams.

\begin{figure}
\begin{center}
\noindent\includegraphics[width=1.0\textwidth]{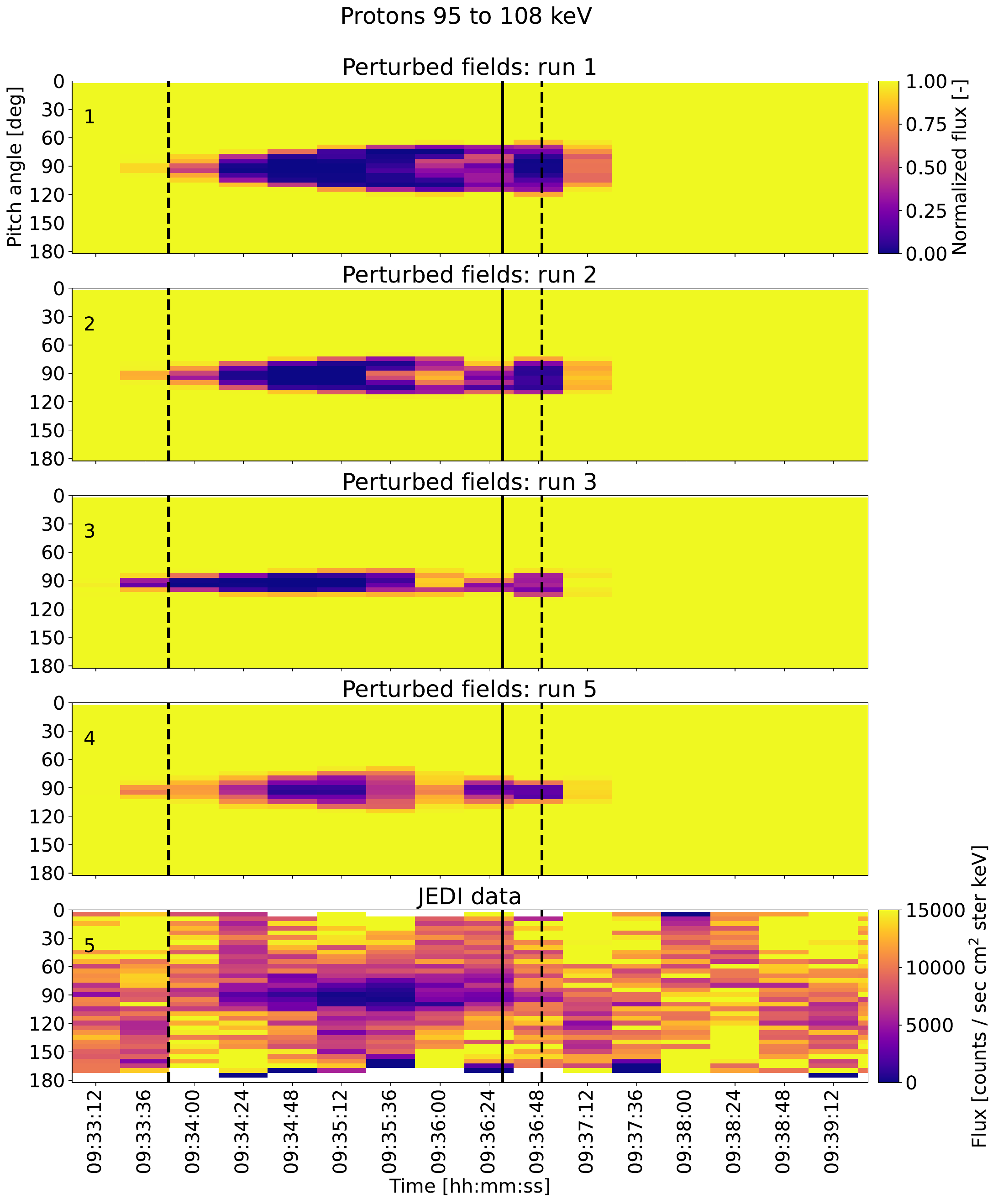}
\end{center}
\caption{Simulations from Figure \ref{fig_sims_fields} (panels 1-4) re-binned in the same time and pitch angle bins as the Juno JEDI data (bottom, panel 5) to allow for a direct comparison. Simulation Run 5 in panel 4 \cite{Cervantes2025} best describes the data, demonstrating the sensitivity of the energetic proton measurements to the precise magnetic field configuration. Note that the colorbar in panel 5 has been saturated at 15000.}
\label{fig_sims_fields_data}
\end{figure}

In Figure \ref{fig_sims_fields_data} the simulations are shown together with the Juno JEDI data. In this case the simulations have been re-binned in the same time and pitch angle bins as the data to allow for a direct comparison. We argue that from a qualitative comparison 'Run 5' describes the JEDI data best. Run 1 predicts a dropout structure that is wider in pitch angle than the data, Run 2 has deeper dropouts near the end of the dropout structure at 09:37:12, while Run 3 is narrower in pitch angle and has a deeper depletion from 09:33:36 to 09:34:00. Finally, Run 5 is the most consistent with the data in terms of depth and extent of the depletion feature. We emphasize that Run 5 best describes both the proton dropout and magnetic field perturbations as observed by Juno \cite{Cervantes2025}. This multi-instrument agreement strengthens the case that the electron beams are a crucial element of Europa's environment.

A quantitative error calculation confirms this conclusion. We calculated the sum of the square of the differences between data and model for each energy and time bin: $\epsilon_{tot} =\sum_{n=1}^{n_{tot}} d_{E_n, \alpha_n}$, in which $d$ is the difference between data and simulation as a function of energy bin $E_n$ and pitch angle bin $\alpha_n$, $n$ is the bin number and $n_{tot}$ the total bin number. Resulting in the following values for $\epsilon_{tot}$: 84.63 (Run 1), 80.29 (Run 2), 82.02 (Run 3) and 77.66 (Run 5). For this computation the data has been normalized using the mean of the time bins from 09:31:12 to 09:33:12, i.e. the data just before the start of the depletion feature. Assuming a simulation with no depletions (i.e. normalized flux equal to 1 everywhere) the error is 95. This indicates that all models improve the fit but that a systematic error likely remains. Possible causes of this error are factors that are not accounted for in the model, including: the variation of the proton distribution in time, the pitch angle distribution of the protons and noise in the measurement. Lastly, we observe that Run 2 has a lower error than Run 3, despite Run 2's lower physical complexity. This suggests that descriptions of the environment might not be mutually exclusive. However, we also caution against using this error calculation as a sole measure of 'correctness' of the simulation, as some differences $d_{E_n, \alpha_n}$ might be more sensitive to the environmental configuration than others. A wide parameter search could help quantify the sensitivity of the dropouts to different environmental configurations and help quantify confidence in a specific solution (see also final paragraph in this section).

The sensitivity of the dropout structure to the magnetic field perturbations and the best agreement between Run 5 and the JEDI data indicates that the energetic proton measurements could be used as a validation of the assumed MHD model which accounts for an asymmetrical atmosphere, non-uniform ionization and electron beams, which is independent of the magnetic field measurements. A key difference being that energetic proton measurements are sensitive to the perturbations in the wider environment that the protons have traversed, as opposed to in-situ magnetic field measurements which only provide a constraint of the field along the trajectory.

While we find the best agreement with 'Run 5', it does not provide for a perfect match with the data either. For example, while a local maximum of depletion occurs near 09:35:36 in the data, this maximum appears shifted to 09:35:12 in the simulation. This discrepancy suggests that further improvements in the description of the environment are possible. Furthermore, the strong sensitivity of the proton dropouts to the different field perturbations indicates that reality may still differ to an important extent from what we have assumed. A future effort to improve a description of Europa's environment should consider magnetic field and energetic ion measurements and models in conjunction. Section 4 in \citeA{Cervantes2025} provides some alternative scenarios that could help guide this effort, such as the inclusion of the ionospheric Hall effect and magnetic field perturbations collocated with electron beams that are partially caused by electric currents flowing along them. Future studies should also address the influence of local magnetic field perturbations at Europa on far field signatures in plasma particle data, e.g. \citeA{Herceg2024}.

\subsection{Weak effect of charge exchange on proton losses}
\label{ss_charge_exchange}
Using a combination of Galileo energetic proton measurements and particle tracing, \citeA{Huybrighs2020} demonstrated that charge exchange with Europa's tenuous atmosphere could cause measurable dropouts during close spacecraft flybys. We find that including charge exchange with the atmosphere of Run 5 from \citeA{Cervantes2025} produces only very minor changes in the simulated dropout at 100 keV that cannot be distinguished due to the limited resolution of the data in time and pitch angle. 

\begin{figure}
\begin{center}
\noindent\includegraphics[width=1.0\textwidth]{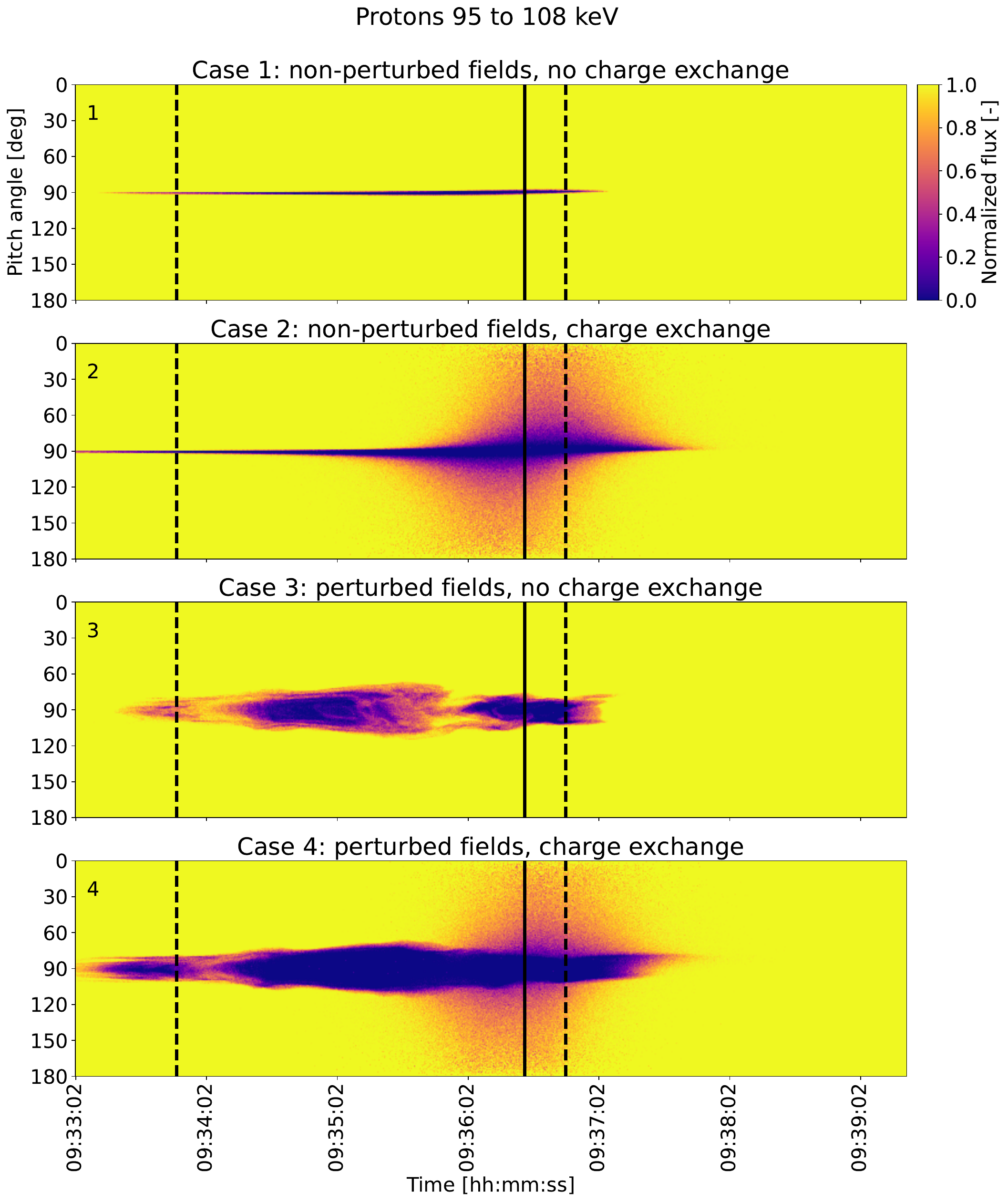}
\end{center}
\caption{In these simulations the hypothetical effect of charge exchange with a dense atmosphere (surface density 10$^{15}$ m$^{-3}$, scale height 100 km) is considered to illustrate the potential effect of an atmosphere with densities well exceeding those in \citeA{Cervantes2025} on the $\sim$ 100 keV proton dropouts in the wake.  The dropout is simulated both without perturbed fields and with perturbed fields (Run 5). We emphasize that we consider the effect of charge exchange on this flyby weak and we see no evidence in the data for the high density case shown in this figure.}
\label{fig_pa_charge_exchange}
\end{figure}

In Figure \ref{fig_pa_charge_exchange} we show losses with a dense atmosphere to get an approximate idea of what a charge exchange dropout would like, if the atmosphere was denser.
The figure shows the effect of charge exchange with a symmetric O$_2$ atmosphere with a surface number density of 10$^{15}$ m$^{-3}$ and a scale height of 100 km. The case with perturbed fields in Figure \ref{fig_pa_charge_exchange} accounts for Run 5 from \citeA{Cervantes2025}, the field perturbations are thus not self-consistent with the assumed atmosphere for the charge exchange. However, we consider this an acceptable first approximation to explore the possible effect of a denser atmosphere. 
We consider this symmetrical atmosphere configuration as an upper limit case in terms of column density and thus of the possible losses due to charge exchange. The assumed density is two orders of magnitude larger than the one used in the MHD runs. Furthermore, the corresponding column density is $10^{20}$ m$^{-2}$ and exceeds the measured typically assumed range of $10^{18}$ to $10^{19}$ m$^{-2}$ (e.g \citeA{Hall1998,Plainaki2018}).

The cases with charge exchange distinguish themselves due to enhanced losses along the full pitch angle range, in particular near the closest approach. We also observe that the loss region along the flyby is more extended. We see no evidence for these extensions in the data and therefore argue that the data are more consistent with Run 5 from \citeA{Cervantes2025}. Thus, even the absence of observed dropouts caused by charged exchange helps constrain the atmospheric properties.
 
The proton dropout is sensitive indirectly to the atmosphere through the perturbed fields (which change for different atmospheric scenarios, e.g. Run 1 and 2) and more directly through charge exchange. Thus, we see the qualitative agreement between our model case Run 5 (Figure \ref{fig_sims_fields_data}) combined with the absence of the extensions of the losses in Figure 
\ref{fig_pa_charge_exchange} as an indirect validation of the chosen atmospheric parameters in Run 5 from \citeA{Cervantes2025}. 

The JUpiter ICy moon Explorer (JUICE) is scheduled to make two flybys of Europa in 2031 and use its instruments to study the moon-magnetosphere interaction at the moon (e.g. \citeA{Masters2025}). Among its instruments is the Particle Environment Package (PEP) which has the capability of detecting energetic ions \cite{Barabash2016}. If JUICE were to encounter higher atmospheric densities, they could cause depletion features that are wide in pitch angle range and extend further along the flyby as in Figure \ref{fig_pa_charge_exchange}.

\subsection{Negligible role of dust on the proton dropouts}
\label{ss_dust}

Using observation of the Juno Waves Instrument, \citeA{Kurth2023}
inferred the number density of dust particles ranging from 0.4 to 3.5$\times10^{-6}$ m$^{-3}$ near Europa during the Juno flyby. We estimate the mean free path (MFP) associated with the dust particles to be approximately $10^{16}$ m.
By comparing these MFP values with the scale of the interaction region, which is at most a few Europa radii, we suggest that interactions of the energetic protons with dust particles are very unlikely, and therefore we neglect any effect of the dust on the proton flux in our simulations. For our calculation we assumed: (1) the mass estimates of the dust from \citeA{Kruger2003} (ranging from 9 to 13$\times 10^{-15}$ kg) obtained through measurements from the Galileo Dust Detector System (DDS) at Europa, (2) spherical dust particles made of water ice (density of 1 g/cm$^3$), and (3) the highest particle number density reported by \citeA{Kurth2023} (3.5$\times10^{-6}$ m$^{-3}$).

\section{Conclusion}
In this study we investigated energetic proton dropouts in Europa's wake measured by the JEDI instrument on the Juno spacecraft, by comparing the measurements to simulations of the proton flux. We show that at energies of $\sim100$ keV the pitch angle morphology of the proton dropout is mostly caused by perturbed electromagnetic fields in Europa's wake. We argue that at $\sim1$ MeV the losses are caused by short half bounce period in combination with the perturbed fields. Furthermore, we show that the dropout morphology at $\sim100$ keV is sensitive to different perturbations of the electromagnetic fields. We find that the MHD model accounting for plasma interaction with an asymmetrical atmosphere, non-uniform ionization and electron beams best describes the measured pitch angle distribution of the dropout. While we find that charge exchange dropouts at $\sim100$ keV energy are weak, this is consistent with the assumed atmospheric properties. We also find the effect of dust in Europa's wake on the proton fluxes are fully  negligible. We conclude that this comparison of proton measurements with particle tracing simulations provides an avenue to validate assumed properties of Europa's environment independent of the magnetic field measurements. Our work emphasizes the importance of energetic ion measurements in the environment of the Galilean moons. After Juno the next opportunity to obtain more such measurements is using the JUICE mission, which is equipped with Particle Environment Package (PEP) that will measure energetic ions during JUICE's two scheduled Europa flybys.

 \section{Open Research}

\begin{itemize}
    \item Particle tracing simulation outputs and simulation code are archived in \citeA{Huybrighs2025_code}. \item Juno JEDI data are available through NASA's Planetary Data System (PDS) \cite{Mauk2024}.
\end{itemize}

\acknowledgments
The work of HH was supported by a DIAS Research Fellowship in Astrophysics. The work of HH, SC, and XC at DIAS was supported by Taighde Éireann - Research Ireland award 22/FFP-P/11545 to CMJ and HH. The work of MH was supported by the the Science Foundation Ireland (Grant 18/FRL/6199) to CMJ, the Discovery programme of the European Space Agency (Contract No: 4000137683/22/NL/GLC/my) and the Research Ireland Pathway Programme (Grant 22/PATH-S/10757). Our gratitude goes to James Peachey and Lawrence E. Brown for their support in using the MIDL software. We also acknowledge the 2024 Dublin Outer Planet Moon‐Magnetosphere Interactions Workshop in fostering collaborations that contributed greatly to this work. We acknowledge the Royal Irish Academy which enabled research visits that supported this study through the 2024-2025 Charlemont Grant Scheme. The content of this publication is solely the responsibility of the authors and does not necessarily represent the official views of the Royal Irish Academy.

\bibliography{references_hh.bib}

\begin{thebibliography}{}

\bibitem [\protect \citeauthoryear {%
Addison%
, Haynes%
, Stahl%
, Liuzzo%
\BCBL {}\ \BBA {} Simon%
}{%
Addison%
\ \protect \BOthers {.}}{%
{\protect \APACyear {2024}}%
}]{%
Addison2024}
\APACinsertmetastar {%
Addison2024}%
\begin{APACrefauthors}%
Addison, P.%
, Haynes, C\BPBI M.%
, Stahl, A\BPBI M.%
, Liuzzo, L.%
\BCBL {}\ \BBA {} Simon, S.%
\end{APACrefauthors}%
\unskip\
\newblock
\APACrefYearMonthDay{2024}{}{}.
\newblock
{\BBOQ}\APACrefatitle {Magnetic Signatures of the Interaction Between Europa and Jupiter's Magnetosphere During the Juno Flyby} {Magnetic signatures of the interaction between europa and jupiter's magnetosphere during the juno flyby}.{\BBCQ}
\newblock
\APACjournalVolNumPages{Geophysical Research Letters}{51}{2}{e2023GL106810}.
\newblock
\begin{APACrefURL} \url{https://agupubs.onlinelibrary.wiley.com/doi/abs/10.1029/2023GL106810} \end{APACrefURL}
\newblock
\APACrefnote{e2023GL106810 2023GL106810}
\newblock
\begin{APACrefDOI} \doi{https://doi.org/10.1029/2023GL106810} \end{APACrefDOI}
\PrintBackRefs{\CurrentBib}

\bibitem [\protect \citeauthoryear {%
Addison%
, Liuzzo%
, Arnold%
\BCBL {}\ \BBA {} Simon%
}{%
Addison%
\ \protect \BOthers {.}}{%
{\protect \APACyear {2021}}%
{\protect \APACexlab {{\protect \BCnt {1}}}}}]{%
Addison2021}
\APACinsertmetastar {%
Addison2021}%
\begin{APACrefauthors}%
Addison, P.%
, Liuzzo, L.%
, Arnold, H.%
\BCBL {}\ \BBA {} Simon, S.%
\end{APACrefauthors}%
\unskip\
\newblock
\APACrefYearMonthDay{2021{\protect \BCnt {1}}}{}{}.
\newblock
{\BBOQ}\APACrefatitle {Influence of Europa’s Time-Varying Electromagnetic Environment on Magnetospheric Ion Precipitation and Surface Weathering} {Influence of europa’s time-varying electromagnetic environment on magnetospheric ion precipitation and surface weathering}.{\BBCQ}
\newblock
\APACjournalVolNumPages{Journal of Geophysical Research: Space Physics}{126}{5}{e2020JA029087}.
\newblock
\begin{APACrefURL} \url{https://agupubs.onlinelibrary.wiley.com/doi/abs/10.1029/2020JA029087} \end{APACrefURL}
\newblock
\APACrefnote{e2020JA029087 2020JA029087}
\newblock
\begin{APACrefDOI} \doi{https://doi.org/10.1029/2020JA029087} \end{APACrefDOI}
\PrintBackRefs{\CurrentBib}

\bibitem [\protect \citeauthoryear {%
Addison%
, Liuzzo%
, Arnold%
\BCBL {}\ \BBA {} Simon%
}{%
Addison%
\ \protect \BOthers {.}}{%
{\protect \APACyear {2021}}%
{\protect \APACexlab {{\protect \BCnt {2}}}}}]{%
Addison2021InfluenceWeathering}
\APACinsertmetastar {%
Addison2021InfluenceWeathering}%
\begin{APACrefauthors}%
Addison, P.%
, Liuzzo, L.%
, Arnold, H.%
\BCBL {}\ \BBA {} Simon, S.%
\end{APACrefauthors}%
\unskip\
\newblock
\APACrefYearMonthDay{2021{\protect \BCnt {2}}}{5}{}.
\newblock
{\BBOQ}\APACrefatitle {{Influence of Europa’s Time-Varying Electromagnetic Environment on Magnetospheric Ion Precipitation and Surface Weathering}} {{Influence of Europa’s Time-Varying Electromagnetic Environment on Magnetospheric Ion Precipitation and Surface Weathering}}.{\BBCQ}
\newblock
\APACjournalVolNumPages{Journal of Geophysical Research: Space Physics}{126}{5}{e2020JA029087}.
\newblock
\begin{APACrefURL} \url{https://onlinelibrary-wiley-com.vu-nl.idm.oclc.org/doi/full/10.1029/2020JA029087 https://onlinelibrary-wiley-com.vu-nl.idm.oclc.org/doi/abs/10.1029/2020JA029087 https://agupubs-onlinelibrary-wiley-com.vu-nl.idm.oclc.org/doi/10.1029/2020JA029087} \end{APACrefURL}
\newblock
\begin{APACrefDOI} \doi{10.1029/2020JA029087} \end{APACrefDOI}
\PrintBackRefs{\CurrentBib}

\bibitem [\protect \citeauthoryear {%
Addison%
, Liuzzo%
\BCBL {}\ \BBA {} Simon%
}{%
Addison%
\ \protect \BOthers {.}}{%
{\protect \APACyear {2022}}%
}]{%
Addison2022}
\APACinsertmetastar {%
Addison2022}%
\begin{APACrefauthors}%
Addison, P.%
, Liuzzo, L.%
\BCBL {}\ \BBA {} Simon, S.%
\end{APACrefauthors}%
\unskip\
\newblock
\APACrefYearMonthDay{2022}{}{}.
\newblock
{\BBOQ}\APACrefatitle {Effect of the Magnetospheric Plasma Interaction and Solar Illumination on Ion Sputtering of Europa’s Surface Ice} {Effect of the magnetospheric plasma interaction and solar illumination on ion sputtering of europa’s surface ice}.{\BBCQ}
\newblock
\APACjournalVolNumPages{Journal of Geophysical Research: Space Physics}{127}{2}{e2021JA030136}.
\newblock
\begin{APACrefURL} \url{https://agupubs.onlinelibrary.wiley.com/doi/abs/10.1029/2021JA030136} \end{APACrefURL}
\newblock
\APACrefnote{e2021JA030136 2021JA030136}
\newblock
\begin{APACrefDOI} \doi{https://doi.org/10.1029/2021JA030136} \end{APACrefDOI}
\PrintBackRefs{\CurrentBib}

\bibitem [\protect \citeauthoryear {%
Allegrini%
\ \protect \BOthers {.}}{%
Allegrini%
\ \protect \BOthers {.}}{%
{\protect \APACyear {2024}}%
}]{%
Allegrini2024}
\APACinsertmetastar {%
Allegrini2024}%
\begin{APACrefauthors}%
Allegrini, F.%
, Saur, J.%
, Szalay, J\BPBI R.%
, Ebert, R\BPBI W.%
, Kurth, W\BPBI S.%
, Cervantes, S.%
\BDBL {}Wilson, R\BPBI J.%
\end{APACrefauthors}%
\unskip\
\newblock
\APACrefYearMonthDay{2024}{}{}.
\newblock
{\BBOQ}\APACrefatitle {Electron Beams at Europa} {Electron beams at europa}.{\BBCQ}
\newblock
\APACjournalVolNumPages{Geophysical Research Letters}{51}{13}{e2024GL108422}.
\newblock
\begin{APACrefURL} \url{https://agupubs.onlinelibrary.wiley.com/doi/abs/10.1029/2024GL108422} \end{APACrefURL}
\newblock
\APACrefnote{e2024GL108422 2024GL108422}
\newblock
\begin{APACrefDOI} \doi{https://doi.org/10.1029/2024GL108422} \end{APACrefDOI}
\PrintBackRefs{\CurrentBib}

\bibitem [\protect \citeauthoryear {%
Barabash%
, Brandt%
, Wurz%
\BCBL {}\ \BBA {} the PEP~team%
}{%
Barabash%
\ \protect \BOthers {.}}{%
{\protect \APACyear {2016}}%
}]{%
Barabash2016}
\APACinsertmetastar {%
Barabash2016}%
\begin{APACrefauthors}%
Barabash, S.%
, Brandt, P\BPBI C.%
, Wurz, P.%
\BCBL {}\ \BBA {} the PEP~team.%
\end{APACrefauthors}%
\unskip\
\newblock
\APACrefYearMonthDay{2016}{}{}.
\newblock
{\BBOQ}\APACrefatitle {Particle Environment Package (PEP) for the ESA JUICE mission} {Particle environment package (pep) for the esa juice mission}.{\BBCQ}
\newblock
\BIn{} \APACrefbtitle {American Astronomical Society, DPS meeting \# 48.} {American astronomical society, dps meeting \# 48.}
\PrintBackRefs{\CurrentBib}

\bibitem [\protect \citeauthoryear {%
Basu%
, Jasperse%
, Robinson%
, Vondrak%
\BCBL {}\ \BBA {} Evans%
}{%
Basu%
\ \protect \BOthers {.}}{%
{\protect \APACyear {1987}}%
}]{%
Basu1987LinearObservations}
\APACinsertmetastar {%
Basu1987LinearObservations}%
\begin{APACrefauthors}%
Basu, B.%
, Jasperse, J\BPBI R.%
, Robinson, R\BPBI M.%
, Vondrak, R\BPBI R.%
\BCBL {}\ \BBA {} Evans, D\BPBI S.%
\end{APACrefauthors}%
\unskip\
\newblock
\APACrefYearMonthDay{1987}{6}{}.
\newblock
{\BBOQ}\APACrefatitle {{Linear transport theory of auroral proton precipitation: A comparison with observations}} {{Linear transport theory of auroral proton precipitation: A comparison with observations}}.{\BBCQ}
\newblock
\APACjournalVolNumPages{Journal of Geophysical Research}{92}{A6}{5920}.
\newblock
\begin{APACrefDOI} \doi{10.1029/ja092ia06p05920} \end{APACrefDOI}
\PrintBackRefs{\CurrentBib}

\bibitem [\protect \citeauthoryear {%
Baumjohann%
\ \BBA {} Treumann%
}{%
Baumjohann%
\ \BBA {} Treumann%
}{%
{\protect \APACyear {1997}}%
}]{%
Baumjohann1997}
\APACinsertmetastar {%
Baumjohann1997}%
\begin{APACrefauthors}%
Baumjohann, W.%
\BCBT {}\ \BBA {} Treumann, R.%
\end{APACrefauthors}%
\unskip\
\newblock
\APACrefYear{1997}.
\newblock
\APACrefbtitle {Basic space plasma physics} {Basic space plasma physics}.
\newblock
\APACaddressPublisher{}{Imperial College Press}.
\PrintBackRefs{\CurrentBib}

\bibitem [\protect \citeauthoryear {%
Breer%
, Liuzzo%
, Arnold%
, Andersson%
\BCBL {}\ \BBA {} Simon%
}{%
Breer%
\ \protect \BOthers {.}}{%
{\protect \APACyear {2019}}%
}]{%
Breer2019}
\APACinsertmetastar {%
Breer2019}%
\begin{APACrefauthors}%
Breer, B\BPBI R.%
, Liuzzo, L.%
, Arnold, H.%
, Andersson, P\BPBI N.%
\BCBL {}\ \BBA {} Simon, S.%
\end{APACrefauthors}%
\unskip\
\newblock
\APACrefYearMonthDay{2019}{}{}.
\newblock
{\BBOQ}\APACrefatitle {Energetic Ion Dynamics in the Perturbed Electromagnetic Fields Near Europa} {Energetic ion dynamics in the perturbed electromagnetic fields near europa}.{\BBCQ}
\newblock
\APACjournalVolNumPages{Journal of Geophysical Research: Space Physics}{124}{9}{7592-7613}.
\newblock
\begin{APACrefURL} \url{https://agupubs.onlinelibrary.wiley.com/doi/abs/10.1029/2019JA027147} \end{APACrefURL}
\newblock
\begin{APACrefDOI} \doi{10.1029/2019JA027147} \end{APACrefDOI}
\PrintBackRefs{\CurrentBib}

\bibitem [\protect \citeauthoryear {%
Cervantes%
\ \protect \BOthers {.}}{%
Cervantes%
\ \protect \BOthers {.}}{%
{\protect \APACyear {2025}}%
}]{%
Cervantes2025}
\APACinsertmetastar {%
Cervantes2025}%
\begin{APACrefauthors}%
Cervantes, S.%
, Saur, J.%
, Duling, S.%
, Szalay, J\BPBI R.%
, Schlegel, S.%
, Connerney, J\BPBI E\BPBI P.%
\BDBL {}Bolton, S.%
\end{APACrefauthors}%
\unskip\
\newblock
\APACrefYearMonthDay{2025}{}{}.
\newblock
{\BBOQ}\APACrefatitle {MHD Simulations of Europa’s Interaction With Jupiter’s Magnetosphere During the Juno Flyby: Electron Beams in the Plasma Wake} {Mhd simulations of europa’s interaction with jupiter’s magnetosphere during the juno flyby: Electron beams in the plasma wake}.{\BBCQ}
\newblock
\APACjournalVolNumPages{Journal of Geophysical Research: Space Physics}{}{}{}.
\newblock
\begin{APACrefURL} \url{https://doi.org/10.22541/essoar.173924045.52275660/v1} \end{APACrefURL}
\newblock
\APACrefnote{In review}
\PrintBackRefs{\CurrentBib}

\bibitem [\protect \citeauthoryear {%
Clark%
, Mauk%
, Paranicas%
, Kollmann%
\BCBL {}\ \BBA {} Smith%
}{%
Clark%
\ \protect \BOthers {.}}{%
{\protect \APACyear {2016}}%
}]{%
Clark2016ChargeMagnetosphere}
\APACinsertmetastar {%
Clark2016ChargeMagnetosphere}%
\begin{APACrefauthors}%
Clark, G.%
, Mauk, B\BPBI H.%
, Paranicas, C.%
, Kollmann, P.%
\BCBL {}\ \BBA {} Smith, H\BPBI T.%
\end{APACrefauthors}%
\unskip\
\newblock
\APACrefYearMonthDay{2016}{3}{}.
\newblock
{\BBOQ}\APACrefatitle {{Charge states of energetic oxygen and sulfur ions in Jupiter's magnetosphere}} {{Charge states of energetic oxygen and sulfur ions in Jupiter's magnetosphere}}.{\BBCQ}
\newblock
\APACjournalVolNumPages{Journal of Geophysical Research: Space Physics}{121}{3}{2264--2273}.
\newblock
\begin{APACrefURL} \url{https://onlinelibrary-wiley-com.vu-nl.idm.oclc.org/doi/full/10.1002/2015JA022257 https://onlinelibrary-wiley-com.vu-nl.idm.oclc.org/doi/abs/10.1002/2015JA022257 https://agupubs-onlinelibrary-wiley-com.vu-nl.idm.oclc.org/doi/10.1002/2015JA022257} \end{APACrefURL}
\newblock
\begin{APACrefDOI} \doi{10.1002/2015JA022257} \end{APACrefDOI}
\PrintBackRefs{\CurrentBib}

\bibitem [\protect \citeauthoryear {%
Clark%
\ \protect \BOthers {.}}{%
Clark%
\ \protect \BOthers {.}}{%
{\protect \APACyear {2025}}%
}]{%
Clark2025}
\APACinsertmetastar {%
Clark2025}%
\begin{APACrefauthors}%
Clark, G.%
, Mauk, B\BPBI H.%
, Paranicas, C\BPBI P.%
, Smith, H\BPBI T.%
, Szalay, J\BPBI R.%
, Kollmann, P.%
\BDBL {}Bolton, S\BPBI J.%
\end{APACrefauthors}%
\unskip\
\newblock
\APACrefYearMonthDay{2025}{}{}.
\newblock
{\BBOQ}\APACrefatitle {Energetic ion losses observed during Juno's close encounter with Europa} {Energetic ion losses observed during juno's close encounter with europa}.{\BBCQ}
\newblock
\APACjournalVolNumPages{Journal of Geophysical Research: Space Physics}{}{}{}.
\newblock
\APACrefnote{In review}
\PrintBackRefs{\CurrentBib}

\bibitem [\protect \citeauthoryear {%
Duling%
\ \protect \BOthers {.}}{%
Duling%
\ \protect \BOthers {.}}{%
{\protect \APACyear {2022}}%
}]{%
Duling2022}
\APACinsertmetastar {%
Duling2022}%
\begin{APACrefauthors}%
Duling, S.%
, Saur, J.%
, Clark, G.%
, Allegrini, F.%
, Greathouse, T.%
, Gladstone, R.%
\BDBL {}Sulaiman, A\BPBI H.%
\end{APACrefauthors}%
\unskip\
\newblock
\APACrefYearMonthDay{2022}{}{}.
\newblock
{\BBOQ}\APACrefatitle {{Ganymede MHD model: Magnetospheric context for Juno's PJ34 flyby}} {{Ganymede MHD model: Magnetospheric context for Juno's PJ34 flyby}}.{\BBCQ}
\newblock
\APACjournalVolNumPages{Geophysical Research Letters}{49}{24}{e2022GL101688}.
\PrintBackRefs{\CurrentBib}

\bibitem [\protect \citeauthoryear {%
Ebert%
\ \protect \BOthers {.}}{%
Ebert%
\ \protect \BOthers {.}}{%
{\protect \APACyear {2025}}%
}]{%
Ebert2025}
\APACinsertmetastar {%
Ebert2025}%
\begin{APACrefauthors}%
Ebert, R\BPBI W.%
, Allegrini, F.%
, Szalay, J\BPBI R.%
, Pontoni, A.%
, Saur, J.%
, Bagenal, F.%
\BDBL {}Wilson, R\BPBI J.%
\end{APACrefauthors}%
\unskip\
\newblock
\APACrefYearMonthDay{2025}{}{}.
\newblock
{\BBOQ}\APACrefatitle {Plasma and Magnetic Field Properties in Europa's Wake From Juno} {Plasma and magnetic field properties in europa's wake from juno}.{\BBCQ}
\newblock
\APACjournalVolNumPages{Journal of Geophysical Research: Space Physics}{130}{5}{e2024JA032520}.
\newblock
\begin{APACrefURL} \url{https://agupubs.onlinelibrary.wiley.com/doi/abs/10.1029/2024JA032520} \end{APACrefURL}
\newblock
\APACrefnote{e2024JA032520 2024JA032520}
\newblock
\begin{APACrefDOI} \doi{https://doi.org/10.1029/2024JA032520} \end{APACrefDOI}
\PrintBackRefs{\CurrentBib}

\bibitem [\protect \citeauthoryear {%
Hall%
, Feldman%
, McGrath%
\BCBL {}\ \BBA {} Strobel%
}{%
Hall%
\ \protect \BOthers {.}}{%
{\protect \APACyear {1998}}%
}]{%
Hall1998}
\APACinsertmetastar {%
Hall1998}%
\begin{APACrefauthors}%
Hall, D.%
, Feldman, P\BPBI D.%
, McGrath, M\BPBI A.%
\BCBL {}\ \BBA {} Strobel, D\BPBI F.%
\end{APACrefauthors}%
\unskip\
\newblock
\APACrefYearMonthDay{1998}{}{}.
\newblock
{\BBOQ}\APACrefatitle {The Far-Ultraviolet Oxygen Airglow of Europa and Ganymede} {The far-ultraviolet oxygen airglow of europa and ganymede}.{\BBCQ}
\newblock
\APACjournalVolNumPages{The Astrophysical Journal}{499}{1}{}.
\newblock
\begin{APACrefDOI} \doi{doi:10.1086/305604} \end{APACrefDOI}
\PrintBackRefs{\CurrentBib}

\bibitem [\protect \citeauthoryear {%
Herceg%
\ \protect \BOthers {.}}{%
Herceg%
\ \protect \BOthers {.}}{%
{\protect \APACyear {2024}}%
}]{%
Herceg2024}
\APACinsertmetastar {%
Herceg2024}%
\begin{APACrefauthors}%
Herceg, M.%
, Jørgensen, J\BPBI L.%
, Denver, T.%
, Jørgensen, P\BPBI S.%
, Benn, M.%
, Connerney, J\BPBI E\BPBI P.%
\BDBL {}Bolton, S\BPBI J.%
\end{APACrefauthors}%
\unskip\
\newblock
\APACrefYearMonthDay{2024}{}{}.
\newblock
{\BBOQ}\APACrefatitle {Europa's Influence on the Jovian Energetic Electron Environment as Observed by Juno's Micro Advanced Stellar Compass} {Europa's influence on the jovian energetic electron environment as observed by juno's micro advanced stellar compass}.{\BBCQ}
\newblock
\APACjournalVolNumPages{Geophysical Research Letters}{51}{9}{e2023GL104685}.
\newblock
\begin{APACrefURL} \url{https://agupubs.onlinelibrary.wiley.com/doi/abs/10.1029/2023GL104685} \end{APACrefURL}
\newblock
\APACrefnote{e2023GL104685 2023GL104685}
\newblock
\begin{APACrefDOI} \doi{https://doi.org/10.1029/2023GL104685} \end{APACrefDOI}
\PrintBackRefs{\CurrentBib}

\bibitem [\protect \citeauthoryear {%
Huybrighs%
}{%
Huybrighs%
}{%
{\protect \APACyear {2025}}%
}]{%
Huybrighs2025_code}
\APACinsertmetastar {%
Huybrighs2025_code}%
\begin{APACrefauthors}%
Huybrighs, H\BPBI L\BPBI F.%
\end{APACrefauthors}%
\unskip\
\newblock
\APACrefYearMonthDay{2025}{}{}.
\newblock
\APACrefbtitle {Particle tracing code, simulation outputs and spacecraft data.} {Particle tracing code, simulation outputs and spacecraft data.}
\newblock
\APAChowpublished {"https://doi.org/10.5281/zenodo.15924107"}.
\PrintBackRefs{\CurrentBib}

\bibitem [\protect \citeauthoryear {%
Huybrighs%
\ \protect \BOthers {.}}{%
Huybrighs%
\ \protect \BOthers {.}}{%
{\protect \APACyear {2023}}%
}]{%
Huybrighs2023}
\APACinsertmetastar {%
Huybrighs2023}%
\begin{APACrefauthors}%
Huybrighs, H\BPBI L\BPBI F.%
, Blöcker, A.%
, Roussos, E.%
, van Buchem, C.%
, Futaana, Y.%
, Holmberg, M\BPBI K\BPBI G.%
\BCBL {}\ \BBA {} Witasse, O.%
\end{APACrefauthors}%
\unskip\
\newblock
\APACrefYearMonthDay{2023}{}{}.
\newblock
{\BBOQ}\APACrefatitle {Europa’s perturbed fields and induced dipole affect energetic proton depletions during distant Alfven wing flybys} {Europa’s perturbed fields and induced dipole affect energetic proton depletions during distant alfven wing flybys}.{\BBCQ}
\newblock
\APACjournalVolNumPages{Geophysical Research Letters}{}{}{}.
\PrintBackRefs{\CurrentBib}

\bibitem [\protect \citeauthoryear {%
Huybrighs%
\ \protect \BOthers {.}}{%
Huybrighs%
\ \protect \BOthers {.}}{%
{\protect \APACyear {2017}}%
}]{%
Huybrighs2017OnMission}
\APACinsertmetastar {%
Huybrighs2017OnMission}%
\begin{APACrefauthors}%
Huybrighs, H\BPBI L\BPBI F.%
, Futaana, Y.%
, Barabash, S.%
, Wieser, M.%
, Wurz, P.%
, Krupp, N.%
\BDBL {}Vermeersen, B.%
\end{APACrefauthors}%
\unskip\
\newblock
\APACrefYearMonthDay{2017}{}{}.
\newblock
{\BBOQ}\APACrefatitle {On the in-situ detectability of Europa's water vapour plumes from a flyby mission} {On the in-situ detectability of europa's water vapour plumes from a flyby mission}.{\BBCQ}
\newblock
\APACjournalVolNumPages{Icarus}{289}{}{270-280}.
\newblock
\begin{APACrefURL} \url{https://www.sciencedirect.com/science/article/pii/S0019103516301968} \end{APACrefURL}
\newblock
\begin{APACrefDOI} \doi{https://doi.org/10.1016/j.icarus.2016.10.026} \end{APACrefDOI}
\PrintBackRefs{\CurrentBib}

\bibitem [\protect \citeauthoryear {%
Huybrighs%
\ \protect \BOthers {.}}{%
Huybrighs%
\ \protect \BOthers {.}}{%
{\protect \APACyear {2020}}%
}]{%
Huybrighs2020}
\APACinsertmetastar {%
Huybrighs2020}%
\begin{APACrefauthors}%
Huybrighs, H\BPBI L\BPBI F.%
, Roussos, E.%
, Blöcker, A.%
, Krupp, N.%
, Futaana, Y.%
, Barabash, S.%
\BDBL {}Witasse, O.%
\end{APACrefauthors}%
\unskip\
\newblock
\APACrefYearMonthDay{2020}{}{}.
\newblock
{\BBOQ}\APACrefatitle {An Active Plume Eruption on Europa During Galileo Flyby E26 as Indicated by Energetic Proton Depletions} {An active plume eruption on europa during galileo flyby e26 as indicated by energetic proton depletions}.{\BBCQ}
\newblock
\APACjournalVolNumPages{Geophysical Research Letters}{47}{10}{e2020GL087806}.
\newblock
\begin{APACrefURL} \url{https://agupubs.onlinelibrary.wiley.com/doi/abs/10.1029/2020GL087806} \end{APACrefURL}
\newblock
\APACrefnote{e2020GL087806 10.1029/2020GL087806}
\newblock
\begin{APACrefDOI} \doi{https://doi.org/10.1029/2020GL087806} \end{APACrefDOI}
\PrintBackRefs{\CurrentBib}

\bibitem [\protect \citeauthoryear {%
Huybrighs%
\ \protect \BOthers {.}}{%
Huybrighs%
\ \protect \BOthers {.}}{%
{\protect \APACyear {2021}}%
}]{%
Huybrighs2021}
\APACinsertmetastar {%
Huybrighs2021}%
\begin{APACrefauthors}%
Huybrighs, H\BPBI L\BPBI F.%
, Roussos, E.%
, Blöcker, A.%
, Krupp, N.%
, Futaana, Y.%
, Barabash, S.%
\BDBL {}Witasse, O.%
\end{APACrefauthors}%
\unskip\
\newblock
\APACrefYearMonthDay{2021}{}{}.
\newblock
{\BBOQ}\APACrefatitle {Reply to Comment on “An Active Plume Eruption on Europa During Galileo Flyby E26 as Indicated by Energetic Proton Depletions”} {Reply to comment on “an active plume eruption on europa during galileo flyby e26 as indicated by energetic proton depletions”}.{\BBCQ}
\newblock
\APACjournalVolNumPages{Geophysical Research Letters}{48}{18}{e2021GL095240}.
\newblock
\begin{APACrefURL} \url{https://agupubs.onlinelibrary.wiley.com/doi/abs/10.1029/2021GL095240} \end{APACrefURL}
\newblock
\APACrefnote{e2021GL095240 2021GL095240}
\newblock
\begin{APACrefDOI} \doi{https://doi.org/10.1029/2021GL095240} \end{APACrefDOI}
\PrintBackRefs{\CurrentBib}

\bibitem [\protect \citeauthoryear {%
Huybrighs%
\ \protect \BOthers {.}}{%
Huybrighs%
\ \protect \BOthers {.}}{%
{\protect \APACyear {2024}}%
}]{%
Huybrighs2024}
\APACinsertmetastar {%
Huybrighs2024}%
\begin{APACrefauthors}%
Huybrighs, H\BPBI L\BPBI F.%
, van Buchem, C\BPBI P\BPBI A.%
, Blöcker, A.%
, Dols, V.%
, Bowers, C\BPBI F.%
\BCBL {}\ \BBA {} Jackman, C\BPBI M.%
\end{APACrefauthors}%
\unskip\
\newblock
\APACrefYearMonthDay{2024}{}{}.
\newblock
{\BBOQ}\APACrefatitle {Energetic Proton Losses Reveal Io's Extended and Longitudinally Asymmetrical Atmosphere} {Energetic proton losses reveal io's extended and longitudinally asymmetrical atmosphere}.{\BBCQ}
\newblock
\APACjournalVolNumPages{Journal of Geophysical Research: Space Physics}{129}{7}{e2023JA032371}.
\newblock
\begin{APACrefURL} \url{https://agupubs.onlinelibrary.wiley.com/doi/abs/10.1029/2023JA032371} \end{APACrefURL}
\newblock
\APACrefnote{e2023JA032371 2023JA032371}
\newblock
\begin{APACrefDOI} \doi{https://doi.org/10.1029/2023JA032371} \end{APACrefDOI}
\PrintBackRefs{\CurrentBib}

\bibitem [\protect \citeauthoryear {%
Kivelson%
, Khurana%
\BCBL {}\ \BBA {} Volwerk%
}{%
Kivelson%
\ \protect \BOthers {.}}{%
{\protect \APACyear {2009}}%
}]{%
Pappalardo2009_Kivelson}
\APACinsertmetastar {%
Pappalardo2009_Kivelson}%
\begin{APACrefauthors}%
Kivelson, M.%
, Khurana, K.%
\BCBL {}\ \BBA {} Volwerk, M.%
\end{APACrefauthors}%
\unskip\
\newblock
\APACrefYearMonthDay{2009}{}{}.
\newblock
{\BBOQ}\APACrefatitle {Europa's Interaction with the Jovian Magnetosphere} {Europa's interaction with the jovian magnetosphere}.{\BBCQ}
\newblock
\BIn{} R\BPBI T.~Pappalardo, W\BPBI B.~McKinnon\BCBL {}\ \BBA {} K.~Khurana\ (\BEDS), \APACrefbtitle {Europa} {Europa}\ (\BPG~545-570).
\newblock
\APACaddressPublisher{}{University of Arizona Press}.
\PrintBackRefs{\CurrentBib}

\bibitem [\protect \citeauthoryear {%
Kotova%
, Roussos%
, Krupp%
\BCBL {}\ \BBA {} Dandouras%
}{%
Kotova%
\ \protect \BOthers {.}}{%
{\protect \APACyear {2015}}%
}]{%
Kotova2015ModelingDione}
\APACinsertmetastar {%
Kotova2015ModelingDione}%
\begin{APACrefauthors}%
Kotova, A.%
, Roussos, E.%
, Krupp, N.%
\BCBL {}\ \BBA {} Dandouras, I.%
\end{APACrefauthors}%
\unskip\
\newblock
\APACrefYearMonthDay{2015}{9}{}.
\newblock
{\BBOQ}\APACrefatitle {{Modeling of the energetic ion observations in the vicinity of Rhea and Dione}} {{Modeling of the energetic ion observations in the vicinity of Rhea and Dione}}.{\BBCQ}
\newblock
\APACjournalVolNumPages{Icarus}{258}{}{402--417}.
\newblock
\begin{APACrefDOI} \doi{10.1016/J.ICARUS.2015.06.031} \end{APACrefDOI}
\PrintBackRefs{\CurrentBib}

\bibitem [\protect \citeauthoryear {%
Krüger%
, Krivov%
, Sremčević%
\BCBL {}\ \BBA {} Grün%
}{%
Krüger%
\ \protect \BOthers {.}}{%
{\protect \APACyear {2003}}%
}]{%
Kruger2003}
\APACinsertmetastar {%
Kruger2003}%
\begin{APACrefauthors}%
Krüger, H.%
, Krivov, A\BPBI V.%
, Sremčević, M.%
\BCBL {}\ \BBA {} Grün, E.%
\end{APACrefauthors}%
\unskip\
\newblock
\APACrefYearMonthDay{2003}{}{}.
\newblock
{\BBOQ}\APACrefatitle {Impact-generated dust clouds surrounding the Galilean moons} {Impact-generated dust clouds surrounding the galilean moons}.{\BBCQ}
\newblock
\APACjournalVolNumPages{Icarus}{164}{1}{170-187}.
\newblock
\begin{APACrefURL} \url{https://www.sciencedirect.com/science/article/pii/S0019103503001271} \end{APACrefURL}
\newblock
\begin{APACrefDOI} \doi{https://doi.org/10.1016/S0019-1035(03)00127-1} \end{APACrefDOI}
\PrintBackRefs{\CurrentBib}

\bibitem [\protect \citeauthoryear {%
Kurth%
\ \protect \BOthers {.}}{%
Kurth%
\ \protect \BOthers {.}}{%
{\protect \APACyear {2023}}%
}]{%
Kurth2023}
\APACinsertmetastar {%
Kurth2023}%
\begin{APACrefauthors}%
Kurth, W\BPBI S.%
, Wilkinson, D\BPBI R.%
, Hospodarsky, G\BPBI B.%
, Santolík, O.%
, Averkamp, T\BPBI F.%
, Sulaiman, A\BPBI H.%
\BDBL {}Bolton, S\BPBI J.%
\end{APACrefauthors}%
\unskip\
\newblock
\APACrefYearMonthDay{2023}{}{}.
\newblock
{\BBOQ}\APACrefatitle {Juno Plasma Wave Observations at Europa} {Juno plasma wave observations at europa}.{\BBCQ}
\newblock
\APACjournalVolNumPages{Geophysical Research Letters}{50}{24}{e2023GL105775}.
\newblock
\begin{APACrefURL} \url{https://agupubs.onlinelibrary.wiley.com/doi/abs/10.1029/2023GL105775} \end{APACrefURL}
\newblock
\APACrefnote{e2023GL105775 2023GL105775}
\newblock
\begin{APACrefDOI} \doi{https://doi.org/10.1029/2023GL105775} \end{APACrefDOI}
\PrintBackRefs{\CurrentBib}

\bibitem [\protect \citeauthoryear {%
Liuzzo%
, Poppe%
, Nénon%
, Simon%
\BCBL {}\ \BBA {} Addison%
}{%
Liuzzo%
\ \protect \BOthers {.}}{%
{\protect \APACyear {2024}}%
}]{%
Liuzzo2024}
\APACinsertmetastar {%
Liuzzo2024}%
\begin{APACrefauthors}%
Liuzzo, L.%
, Poppe, A\BPBI R.%
, Nénon, Q.%
, Simon, S.%
\BCBL {}\ \BBA {} Addison, P.%
\end{APACrefauthors}%
\unskip\
\newblock
\APACrefYearMonthDay{2024}{}{}.
\newblock
{\BBOQ}\APACrefatitle {Constraining the Influence of Callisto's Perturbed Electromagnetic Environment on Energetic Particle Observations} {Constraining the influence of callisto's perturbed electromagnetic environment on energetic particle observations}.{\BBCQ}
\newblock
\APACjournalVolNumPages{Journal of Geophysical Research: Space Physics}{129}{2}{e2023JA032189}.
\newblock
\begin{APACrefURL} \url{https://agupubs.onlinelibrary.wiley.com/doi/abs/10.1029/2023JA032189} \end{APACrefURL}
\newblock
\APACrefnote{e2023JA032189 2023JA032189}
\newblock
\begin{APACrefDOI} \doi{https://doi.org/10.1029/2023JA032189} \end{APACrefDOI}
\PrintBackRefs{\CurrentBib}

\bibitem [\protect \citeauthoryear {%
Liuzzo%
, Simon%
\BCBL {}\ \BBA {} Regoli%
}{%
Liuzzo%
\ \protect \BOthers {.}}{%
{\protect \APACyear {2019}}%
}]{%
Liuzzo2019EnergeticCallistob}
\APACinsertmetastar {%
Liuzzo2019EnergeticCallistob}%
\begin{APACrefauthors}%
Liuzzo, L.%
, Simon, S.%
\BCBL {}\ \BBA {} Regoli, L.%
\end{APACrefauthors}%
\unskip\
\newblock
\APACrefYearMonthDay{2019}{2}{}.
\newblock
{\BBOQ}\APACrefatitle {{Energetic ion dynamics near Callisto}} {{Energetic ion dynamics near Callisto}}.{\BBCQ}
\newblock
\APACjournalVolNumPages{Planetary and Space Science}{166}{}{23--53}.
\newblock
\begin{APACrefDOI} \doi{10.1016/J.PSS.2018.07.014} \end{APACrefDOI}
\PrintBackRefs{\CurrentBib}

\bibitem [\protect \citeauthoryear {%
Masters%
}{%
Masters%
}{%
{\protect \APACyear {2025}}%
}]{%
Masters2025}
\APACinsertmetastar {%
Masters2025}%
\begin{APACrefauthors}%
Masters, A.%
\end{APACrefauthors}%
\unskip\
\newblock
\APACrefYearMonthDay{2025}{}{}.
\newblock
{\BBOQ}\APACrefatitle {{place holder for masters et al., 2025}} {{place holder for masters et al., 2025}}.{\BBCQ}
\newblock
\APACjournalVolNumPages{SSR}{}{}{}.
\PrintBackRefs{\CurrentBib}

\bibitem [\protect \citeauthoryear {%
B.~Mauk%
}{%
B.~Mauk%
}{%
{\protect \APACyear {2024}}%
}]{%
Mauk2024}
\APACinsertmetastar {%
Mauk2024}%
\begin{APACrefauthors}%
Mauk, B.%
\end{APACrefauthors}%
\unskip\
\newblock
\APACrefYearMonthDay{2024}{}{}.
\newblock
\APACrefbtitle {{JEDI CALIBRATED (CDR) DATA JNO J JED 3 CDR V1.0}.} {{JEDI CALIBRATED (CDR) DATA JNO J JED 3 CDR V1.0}.}
\newblock
\APACaddressPublisher{}{NASA Planetary Data System}.
\newblock
\begin{APACrefDOI} \doi{10.17189/1519713} \end{APACrefDOI}
\PrintBackRefs{\CurrentBib}

\bibitem [\protect \citeauthoryear {%
B\BPBI H.~Mauk%
\ \protect \BOthers {.}}{%
B\BPBI H.~Mauk%
\ \protect \BOthers {.}}{%
{\protect \APACyear {2017}}%
}]{%
Mauk2017TheMission}
\APACinsertmetastar {%
Mauk2017TheMission}%
\begin{APACrefauthors}%
Mauk, B\BPBI H.%
, Haggerty, D\BPBI K.%
, Jaskulek, S\BPBI E.%
, Schlemm, C\BPBI E.%
, Brown, L\BPBI E.%
, Cooper, S\BPBI A.%
\BDBL {}Stokes, M\BPBI R.%
\end{APACrefauthors}%
\unskip\
\newblock
\APACrefYearMonthDay{2017}{11}{}.
\newblock
{\BBOQ}\APACrefatitle {{The Jupiter Energetic Particle Detector Instrument (JEDI) Investigation for the Juno Mission}} {{The Jupiter Energetic Particle Detector Instrument (JEDI) Investigation for the Juno Mission}}.{\BBCQ}
\newblock
\APACjournalVolNumPages{Space Science Reviews}{213}{1-4}{289--346}.
\newblock
\begin{APACrefURL} \url{https://link-springer-com.vu-nl.idm.oclc.org/article/10.1007/s11214-013-0025-3} \end{APACrefURL}
\newblock
\begin{APACrefDOI} \doi{10.1007/S11214-013-0025-3/TABLES/21} \end{APACrefDOI}
\PrintBackRefs{\CurrentBib}

\bibitem [\protect \citeauthoryear {%
Mignone%
\ \protect \BOthers {.}}{%
Mignone%
\ \protect \BOthers {.}}{%
{\protect \APACyear {2007}}%
}]{%
Mignone2007}
\APACinsertmetastar {%
Mignone2007}%
\begin{APACrefauthors}%
Mignone, A.%
, Bodo, G.%
, Massaglia, S.%
, Matsakos, T.%
, Tesileanu, O\BPBI e.%
, Zanni, C.%
\BCBL {}\ \BBA {} Ferrari, A.%
\end{APACrefauthors}%
\unskip\
\newblock
\APACrefYearMonthDay{2007}{}{}.
\newblock
{\BBOQ}\APACrefatitle {{PLUTO: a numerical code for computational astrophysics}} {{PLUTO: a numerical code for computational astrophysics}}.{\BBCQ}
\newblock
\APACjournalVolNumPages{The Astrophysical Journal Supplement Series}{170}{1}{228}.
\PrintBackRefs{\CurrentBib}

\bibitem [\protect \citeauthoryear {%
Nordheim%
\ \protect \BOthers {.}}{%
Nordheim%
\ \protect \BOthers {.}}{%
{\protect \APACyear {2022}}%
}]{%
Nordheim2022}
\APACinsertmetastar {%
Nordheim2022}%
\begin{APACrefauthors}%
Nordheim, T\BPBI A.%
, Regoli, L\BPBI H.%
, Harris, C\BPBI D.%
, Paranicas, C.%
, Hand, K\BPBI P.%
\BCBL {}\ \BBA {} Jia, X.%
\end{APACrefauthors}%
\unskip\
\newblock
\APACrefYearMonthDay{2022}{1}{}.
\newblock
{\BBOQ}\APACrefatitle {{Magnetospheric Ion Bombardment of Europa’s Surface}} {{Magnetospheric Ion Bombardment of Europa’s Surface}}.{\BBCQ}
\newblock
\APACjournalVolNumPages{The Planetary Science Journal}{3}{1}{5}.
\newblock
\begin{APACrefURL} \url{https://iopscience-iop-org.vu-nl.idm.oclc.org/article/10.3847/PSJ/ac382a https://iopscience-iop-org.vu-nl.idm.oclc.org/article/10.3847/PSJ/ac382a/meta} \end{APACrefURL}
\newblock
\begin{APACrefDOI} \doi{10.3847/PSJ/AC382A} \end{APACrefDOI}
\PrintBackRefs{\CurrentBib}

\bibitem [\protect \citeauthoryear {%
Paranicas%
\ \protect \BOthers {.}}{%
Paranicas%
\ \protect \BOthers {.}}{%
{\protect \APACyear {2023}}%
}]{%
Paranicas2023}
\APACinsertmetastar {%
Paranicas2023}%
\begin{APACrefauthors}%
Paranicas, C.%
, Mauk, B\BPBI H.%
, Clark, G.%
, Kollmann, P.%
, Westlake, J.%
, Hibbitts, K.%
\BDBL {}Bolton, S.%
\end{APACrefauthors}%
\unskip\
\newblock
\APACrefYearMonthDay{2023}{}{}.
\newblock
{\BBOQ}\APACrefatitle {Energetic Electrons Near Europa From Juno JEDI Data} {Energetic electrons near europa from juno jedi data}.{\BBCQ}
\newblock
\APACjournalVolNumPages{Geophysical Research Letters}{50}{21}{e2023GL105598}.
\newblock
\begin{APACrefURL} \url{https://agupubs.onlinelibrary.wiley.com/doi/abs/10.1029/2023GL105598} \end{APACrefURL}
\newblock
\APACrefnote{e2023GL105598 2023GL105598}
\newblock
\begin{APACrefDOI} \doi{https://doi.org/10.1029/2023GL105598} \end{APACrefDOI}
\PrintBackRefs{\CurrentBib}

\bibitem [\protect \citeauthoryear {%
Paranicas%
\ \protect \BOthers {.}}{%
Paranicas%
\ \protect \BOthers {.}}{%
{\protect \APACyear {2007}}%
}]{%
Paranicas2007}
\APACinsertmetastar {%
Paranicas2007}%
\begin{APACrefauthors}%
Paranicas, C.%
, Mauk, B\BPBI H.%
, Khurana, K.%
, Jun, I.%
, Garrett, H.%
, Krupp, N.%
\BCBL {}\ \BBA {} Roussos, E.%
\end{APACrefauthors}%
\unskip\
\newblock
\APACrefYearMonthDay{2007}{}{}.
\newblock
{\BBOQ}\APACrefatitle {Europa's near-surface radiation environment} {Europa's near-surface radiation environment}.{\BBCQ}
\newblock
\APACjournalVolNumPages{Geophysical Research Letters}{34}{15}{}.
\newblock
\begin{APACrefURL} \url{https://agupubs.onlinelibrary.wiley.com/doi/abs/10.1029/2007GL030834} \end{APACrefURL}
\newblock
\begin{APACrefDOI} \doi{10.1029/2007GL030834} \end{APACrefDOI}
\PrintBackRefs{\CurrentBib}

\bibitem [\protect \citeauthoryear {%
Paranicas%
, McEntire%
, Cheng%
, Lagg%
\BCBL {}\ \BBA {} Williams%
}{%
Paranicas%
\ \protect \BOthers {.}}{%
{\protect \APACyear {2000}}%
{\protect \APACexlab {{\protect \BCnt {1}}}}}]{%
Paranicas2000}
\APACinsertmetastar {%
Paranicas2000}%
\begin{APACrefauthors}%
Paranicas, C.%
, McEntire, R\BPBI W.%
, Cheng, A\BPBI F.%
, Lagg, A.%
\BCBL {}\ \BBA {} Williams, D\BPBI J.%
\end{APACrefauthors}%
\unskip\
\newblock
\APACrefYearMonthDay{2000{\protect \BCnt {1}}}{}{}.
\newblock
{\BBOQ}\APACrefatitle {Energetic charged particles near Europa} {Energetic charged particles near europa}.{\BBCQ}
\newblock
\APACjournalVolNumPages{Journal of Geophysical Research: Space Physics}{105}{A7}{16005--16015}.
\newblock
\begin{APACrefDOI} \doi{10.1029/1999JA000350} \end{APACrefDOI}
\PrintBackRefs{\CurrentBib}

\bibitem [\protect \citeauthoryear {%
Paranicas%
, McEntire%
, Cheng%
, Lagg%
\BCBL {}\ \BBA {} Williams%
}{%
Paranicas%
\ \protect \BOthers {.}}{%
{\protect \APACyear {2000}}%
{\protect \APACexlab {{\protect \BCnt {2}}}}}]{%
Paranicas2000EnergeticEuropa}
\APACinsertmetastar {%
Paranicas2000EnergeticEuropa}%
\begin{APACrefauthors}%
Paranicas, C.%
, McEntire, R\BPBI W.%
, Cheng, A\BPBI F.%
, Lagg, A.%
\BCBL {}\ \BBA {} Williams, D\BPBI J.%
\end{APACrefauthors}%
\unskip\
\newblock
\APACrefYearMonthDay{2000{\protect \BCnt {2}}}{7}{}.
\newblock
{\BBOQ}\APACrefatitle {{Energetic charged particles near Europa}} {{Energetic charged particles near Europa}}.{\BBCQ}
\newblock
\APACjournalVolNumPages{Journal of Geophysical Research: Space Physics}{105}{A7}{16005--16015}.
\newblock
\begin{APACrefURL} \url{https://agupubs.onlinelibrary.wiley.com/doi/full/10.1029/1999JA000350 https://agupubs.onlinelibrary.wiley.com/doi/abs/10.1029/1999JA000350 https://agupubs.onlinelibrary.wiley.com/doi/10.1029/1999JA000350} \end{APACrefURL}
\newblock
\begin{APACrefDOI} \doi{10.1029/1999ja000350} \end{APACrefDOI}
\PrintBackRefs{\CurrentBib}

\bibitem [\protect \citeauthoryear {%
Parisi%
\ \protect \BOthers {.}}{%
Parisi%
\ \protect \BOthers {.}}{%
{\protect \APACyear {2023}}%
}]{%
Parisi2023}
\APACinsertmetastar {%
Parisi2023}%
\begin{APACrefauthors}%
Parisi, M.%
, Caruso, A.%
, Buccino, D\BPBI R.%
, Gramigna, E.%
, Withers, P.%
, Gomez-Casajus, L.%
\BDBL {}Bolton, S.%
\end{APACrefauthors}%
\unskip\
\newblock
\APACrefYearMonthDay{2023}{}{}.
\newblock
{\BBOQ}\APACrefatitle {Radio Occultation Measurements of Europa's Ionosphere From Juno's Close Flyby} {Radio occultation measurements of europa's ionosphere from juno's close flyby}.{\BBCQ}
\newblock
\APACjournalVolNumPages{Geophysical Research Letters}{50}{22}{e2023GL106637}.
\newblock
\begin{APACrefURL} \url{https://agupubs.onlinelibrary.wiley.com/doi/abs/10.1029/2023GL106637} \end{APACrefURL}
\newblock
\APACrefnote{e2023GL106637 2023GL106637}
\newblock
\begin{APACrefDOI} \doi{https://doi.org/10.1029/2023GL106637} \end{APACrefDOI}
\PrintBackRefs{\CurrentBib}

\bibitem [\protect \citeauthoryear {%
Plainaki%
\ \protect \BOthers {.}}{%
Plainaki%
\ \protect \BOthers {.}}{%
{\protect \APACyear {2018}}%
}]{%
Plainaki2018}
\APACinsertmetastar {%
Plainaki2018}%
\begin{APACrefauthors}%
Plainaki, C.%
, Cassidy, T\BPBI A.%
, Shematovich, V\BPBI I.%
, Milillo, A.%
, Wurz, P.%
, Vorburger, A.%
\BDBL {}Teolis, B.%
\end{APACrefauthors}%
\unskip\
\newblock
\APACrefYearMonthDay{2018}{Jan}{24}.
\newblock
{\BBOQ}\APACrefatitle {Towards a Global Unified Model of Europa's Tenuous Atmosphere} {Towards a global unified model of europa's tenuous atmosphere}.{\BBCQ}
\newblock
\APACjournalVolNumPages{Space Science Reviews}{214}{1}{40}.
\newblock
\begin{APACrefURL} \url{https://doi.org/10.1007/s11214-018-0469-6} \end{APACrefURL}
\newblock
\begin{APACrefDOI} \doi{10.1007/s11214-018-0469-6} \end{APACrefDOI}
\PrintBackRefs{\CurrentBib}

\bibitem [\protect \citeauthoryear {%
Plainaki%
\ \protect \BOthers {.}}{%
Plainaki%
\ \protect \BOthers {.}}{%
{\protect \APACyear {2020}}%
}]{%
Plainaki2020}
\APACinsertmetastar {%
Plainaki2020}%
\begin{APACrefauthors}%
Plainaki, C.%
, Massetti, S.%
, Jia, X.%
, Mura, A.%
, Milillo, A.%
, Grassi, D.%
\BDBL {}Filacchione, G.%
\end{APACrefauthors}%
\unskip\
\newblock
\APACrefYearMonthDay{2020}{sep}{}.
\newblock
{\BBOQ}\APACrefatitle {Kinetic Simulations of the Jovian Energetic Ion Circulation around Ganymede} {Kinetic simulations of the jovian energetic ion circulation around ganymede}.{\BBCQ}
\newblock
\APACjournalVolNumPages{The Astrophysical Journal}{900}{1}{74}.
\newblock
\begin{APACrefURL} \url{https://doi.org/10.3847/1538-4357/aba94c} \end{APACrefURL}
\newblock
\begin{APACrefDOI} \doi{10.3847/1538-4357/aba94c} \end{APACrefDOI}
\PrintBackRefs{\CurrentBib}

\bibitem [\protect \citeauthoryear {%
Poppe%
, Fatemi%
\BCBL {}\ \BBA {} Khurana%
}{%
Poppe%
\ \protect \BOthers {.}}{%
{\protect \APACyear {2018}}%
}]{%
Poppe2018ThermalMagnetosphere}
\APACinsertmetastar {%
Poppe2018ThermalMagnetosphere}%
\begin{APACrefauthors}%
Poppe, A\BPBI R.%
, Fatemi, S.%
\BCBL {}\ \BBA {} Khurana, K\BPBI K.%
\end{APACrefauthors}%
\unskip\
\newblock
\APACrefYearMonthDay{2018}{6}{}.
\newblock
{\BBOQ}\APACrefatitle {{Thermal and Energetic Ion Dynamics in Ganymede's Magnetosphere}} {{Thermal and Energetic Ion Dynamics in Ganymede's Magnetosphere}}.{\BBCQ}
\newblock
\APACjournalVolNumPages{Journal of Geophysical Research: Space Physics}{123}{6}{4614--4637}.
\newblock
\begin{APACrefURL} \url{https://onlinelibrary-wiley-com.vu-nl.idm.oclc.org/doi/full/10.1029/2018JA025312 https://onlinelibrary-wiley-com.vu-nl.idm.oclc.org/doi/abs/10.1029/2018JA025312 https://agupubs-onlinelibrary-wiley-com.vu-nl.idm.oclc.org/doi/10.1029/2018JA025312} \end{APACrefURL}
\newblock
\begin{APACrefDOI} \doi{10.1029/2018JA025312} \end{APACrefDOI}
\PrintBackRefs{\CurrentBib}

\bibitem [\protect \citeauthoryear {%
Saur%
, Strobel%
\BCBL {}\ \BBA {} Neubauer%
}{%
Saur%
\ \protect \BOthers {.}}{%
{\protect \APACyear {1998}}%
}]{%
Saur1998InteractionAtmosphere}
\APACinsertmetastar {%
Saur1998InteractionAtmosphere}%
\begin{APACrefauthors}%
Saur, J.%
, Strobel, D\BPBI F.%
\BCBL {}\ \BBA {} Neubauer, F\BPBI M.%
\end{APACrefauthors}%
\unskip\
\newblock
\APACrefYearMonthDay{1998}{8}{}.
\newblock
{\BBOQ}\APACrefatitle {{Interaction of the Jovian magnetosphere with Europa: Constraints on the neutral atmosphere}} {{Interaction of the Jovian magnetosphere with Europa: Constraints on the neutral atmosphere}}.{\BBCQ}
\newblock
\APACjournalVolNumPages{Journal of Geophysical Research E: Planets}{103}{3339}{19947--19962}.
\newblock
\begin{APACrefDOI} \doi{10.1029/97je03556} \end{APACrefDOI}
\PrintBackRefs{\CurrentBib}

\bibitem [\protect \citeauthoryear {%
Selesnick%
\ \BBA {} Cohen%
}{%
Selesnick%
\ \BBA {} Cohen%
}{%
{\protect \APACyear {2009}}%
}]{%
Selesnick2009ChargeIo}
\APACinsertmetastar {%
Selesnick2009ChargeIo}%
\begin{APACrefauthors}%
Selesnick, R\BPBI S.%
\BCBT {}\ \BBA {} Cohen, C\BPBI M.%
\end{APACrefauthors}%
\unskip\
\newblock
\APACrefYearMonthDay{2009}{1}{}.
\newblock
{\BBOQ}\APACrefatitle {{Charge states of energetic ions in Jupiter's radiation belt inferred from absorption microsignatures of Io}} {{Charge states of energetic ions in Jupiter's radiation belt inferred from absorption microsignatures of Io}}.{\BBCQ}
\newblock
\APACjournalVolNumPages{Journal of Geophysical Research: Space Physics}{114}{A1}{1207}.
\newblock
\begin{APACrefURL} \url{https://onlinelibrary-wiley-com.vu-nl.idm.oclc.org/doi/full/10.1029/2008JA013722 https://onlinelibrary-wiley-com.vu-nl.idm.oclc.org/doi/abs/10.1029/2008JA013722 https://agupubs-onlinelibrary-wiley-com.vu-nl.idm.oclc.org/doi/10.1029/2008JA013722} \end{APACrefURL}
\newblock
\begin{APACrefDOI} \doi{10.1029/2008JA013722} \end{APACrefDOI}
\PrintBackRefs{\CurrentBib}

\bibitem [\protect \citeauthoryear {%
Strack%
\ \BBA {} Saur%
}{%
Strack%
\ \BBA {} Saur%
}{%
{\protect \APACyear {2024}}%
}]{%
Strack2024}
\APACinsertmetastar {%
Strack2024}%
\begin{APACrefauthors}%
Strack, D.%
\BCBT {}\ \BBA {} Saur, J.%
\end{APACrefauthors}%
\unskip\
\newblock
\APACrefYearMonthDay{2024}{}{}.
\newblock
{\BBOQ}\APACrefatitle {{The spatiotemporal structure of induced magnetic fields in Callisto's plasma environment due to their propagation with MHD modes}} {{The spatiotemporal structure of induced magnetic fields in Callisto's plasma environment due to their propagation with MHD modes}}.{\BBCQ}
\newblock
\APACjournalVolNumPages{Journal of Geophysical Research: Space Physics}{129}{12}{e2024JA033235}.
\PrintBackRefs{\CurrentBib}

\bibitem [\protect \citeauthoryear {%
Szalay%
\ \protect \BOthers {.}}{%
Szalay%
\ \protect \BOthers {.}}{%
{\protect \APACyear {2024}}%
}]{%
Szalay2024}
\APACinsertmetastar {%
Szalay2024}%
\begin{APACrefauthors}%
Szalay, J\BPBI R.%
, Allegrini, F.%
, Ebert, R\BPBI W.%
, Bagenal, F.%
, Bolton, S\BPBI J.%
, Fatemi, S.%
\BDBL {}Wilson, R\BPBI J.%
\end{APACrefauthors}%
\unskip\
\newblock
\APACrefYearMonthDay{2024}{May}{01}.
\newblock
{\BBOQ}\APACrefatitle {Oxygen production from dissociation of Europa's water-ice surface} {Oxygen production from dissociation of europa's water-ice surface}.{\BBCQ}
\newblock
\APACjournalVolNumPages{Nature Astronomy}{8}{5}{567-576}.
\newblock
\begin{APACrefURL} \url{https://doi.org/10.1038/s41550-024-02206-x} \end{APACrefURL}
\newblock
\begin{APACrefDOI} \doi{10.1038/s41550-024-02206-x} \end{APACrefDOI}
\PrintBackRefs{\CurrentBib}

\end{thebibliography}

\end{document}